\documentclass{emulateapj}
\usepackage{color}
\usepackage{amsmath}  
\usepackage[colorlinks=true,citecolor=magenta]{hyperref}

\newcommand{\Mpc}{\mathrm{Mpc}}
\newcommand{\yr}{\mathrm{yr}}
\newcommand{\xp}{{\sc X-Pipeline}}
\newcommand{\vo}{{Virgo}}  
\newcommand{\lo}{{\sc LIGO}}
\newcommand{\ant}{{\sc ANTARES}}
\newcommand{\ice}{{\sc IceCube}}
\newcommand{\km}{{\sc KM3NeT}}
\newcommand{\dccnumber}{P1200006}

\begin{document}

\title{A First Search for coincident Gravitational Waves and High Energy Neutrinos using LIGO, \vo~and \ant~data from 2007}

\author{S.~Adri\'an-Mart\'inez\altaffilmark{1}, 
I.~Al Samarai\altaffilmark{2},  
A.~Albert\altaffilmark{3}, 
M.~Andr\'e\altaffilmark{4},  
M.~Anghinolfi\altaffilmark{5},  
G.~Anton\altaffilmark{6},
S.~Anvar\altaffilmark{7}, 
M.~Ardid\altaffilmark{1}, 
T.~Astraatmadja\altaffilmark{8,39},
J-J.~Aubert\altaffilmark{2}
% ...
}
\affil{The {\sc Antares} Collaboration}

\author{J.~Aasi\altaffilmark{42},
J.~Abadie\altaffilmark{42},
B.~P.~Abbott\altaffilmark{42},
R.~Abbott\altaffilmark{42},
T.~D.~Abbott\altaffilmark{43},
M.~Abernathy\altaffilmark{44},
T.~Accadia\altaffilmark{45},
F.~Acernese\altaffilmark{46,48},
C.~Adams\altaffilmark{49},
T.~Adams\altaffilmark{50}
% ...
}
\affil{The {\sc LIGO} Scientific Collaboration and the {\sc Virgo} Collaboration}

\altaffiltext{1}{\label{ant1}Institut d'Investigaci\'o per a la Gesti\'o Integrada de les Zones Costaneres (IGIC) - Universitat Polit\`ecnica de Val\`encia. C/  Paranimf 1 , 46730 Gandia, Spain.}
\altaffiltext{2}{\label{ant3}CPPM, Aix-Marseille Universit\'e, CNRS/IN2P3, Marseille, France}
\altaffiltext{3}{\label{ant4}GRPHE - Institut universitaire de technologie de Colmar, 34 rue du Grillenbreit BP 50568 - 68008 Colmar, France }
\altaffiltext{4}{\label{ant5}Technical University of Catalonia, Laboratory of Applied Bioacoustics, Rambla Exposici\'o, 08800 Vilanova i la Geltr\'u, Barcelona, Spain}
\altaffiltext{5}{\label{ant6}INFN - Sezione di Genova, Via Dodecaneso 33, 16146 Genova, Italy}
\altaffiltext{6}{\label{ant7}Friedrich-Alexander-Universit\"at Erlangen-N\"urnberg, Erlangen Centre for Astroparticle Physics, Erwin-Rommel-Str. 1, 91058 Erlangen, Germany}
\altaffiltext{7}{\label{ant8}Direction des Sciences de la Mati\`ere - Institut de recherche sur les lois fondamentales de l'Univers - Service d'Electronique des D\'etecteurs et d'Informatique, CEA Saclay, 91191 Gif-sur-Yvette Cedex, France}
\altaffiltext{8}{\label{ant9}Nikhef, Science Park,  Amsterdam, The Netherlands}
% ...
\altaffiltext{39}{\label{anta} Also at University of Leiden, the Netherlands} 
% ...
\altaffiltext{42}{\label{lvc1}LIGO - California Institute of Technology, Pasadena, CA  91125, USA }
\altaffiltext{43}{\label{lvc2}California State University Fullerton, Fullerton CA 92831 USA}
\altaffiltext{44}{\label{lvc3}SUPA, University of Glasgow, Glasgow, G12 8QQ, United Kingdom }
\altaffiltext{45}{\label{lvc4}Laboratoire d'Annecy-le-Vieux de Physique des Particules (LAPP), Universit\'e de Savoie, CNRS/IN2P3, F-74941 Annecy-Le-Vieux, France}
\altaffiltext{46}{\label{lvc5a}INFN, Sezione di Napoli, Complesso Universitario di Monte S.Angelo, I-80126 Napoli, Italy}
% ...
\altaffiltext{48}{\label{lvc5c}Universit\`a di Salerno, I-84084 Fisciano (Salerno), Italy}
\altaffiltext{49}{\label{lvc6}LIGO - Livingston Observatory, Livingston, LA  70754, USA }
\altaffiltext{50}{\label{lvc7}Cardiff University, Cardiff, CF24 3AA, United Kingdom }
% ...

\shorttitle{Coincidences between Gravitational Waves and High Energy Neutrinos}
\shortauthors{The \ant~Collaboration, the \lo~Scientific Collaboration and the \vo~Collaboration}

\begin{abstract}

We present the results of the first search for gravitational wave bursts 
associated with high energy neutrinos.  
Together, these messengers could reveal 
new, hidden sources that are not observed by conventional photon astronomy, 
particularly at high energy.  Our search uses neutrinos detected by the 
underwater neutrino telescope \ant~in its 5 line configuration during the 
period January - September 2007, which coincided with the fifth and first science runs of \lo~and \vo,
respectively.  The {\lo}-{\vo} data were analysed for 
candidate gravitational-wave signals coincident in time and direction with 
the neutrino events.
No significant coincident events were observed.  We place limits on the 
density of joint high energy neutrino - gravitational wave 
emission events in the local universe, and compare them with densities of merger and core-collapse events.
\end{abstract}

\shorttitle{Coincidences between Gravitational Waves and High Energy Neutrinos}
\shortauthors{The \ant~Collaboration, the \lo~Scientific Collaboration and the \vo~Collaboration}

\section{Introduction}

Multi-messenger astronomy is entering a stimulating period with the
recent development of experimental techniques that will open new
windows of cosmic radiation observation in all its components.  In
particular, both high-energy ($\gg$\,GeV) neutrinos (HENs) and
gravitational waves (GWs), neither of which have yet been directly
observed from astrophysical sources, are becoming new tools for
exploring the Universe.

While HENs are expected to be produced in interactions between
relativistic protons and the external radiation field of the source
(e.g., \citealt{waxman:97,huemmer:12}), GWs carry information on the
intricate multi-dimensional dynamics in the source's central regions
(e.g., \citealt{creighton:11}). HENs and GWs are thus complementary
messengers.

Simultaneous emission of GWs and HENs has been proposed in a
range of cataclysmic cosmic events including gamma-ray bursts (GRBs),
core-collapse supernovae (CCSNe), soft-gamma repeater outbursts
(SGRs), and, potentially, cosmic string cusps in the early universe.

Observational constraints on HEN and GW emission from some of these
phenomena have already been obtained. The IceCube collaboration
recently placed limits on the HEN emission in GRBs
(\citealt{abbasi:12grb,abbasi:11grb,abbasi:10grb}), SGRs and blazars
\citep{abbasi:12allsky}, and jet-driven CCSNe \citep{abbasi:12ccsn}
using data from the IceCube detector at various levels of
completion. Similarly, the \ant~Collaboration has placed limits on the
HEN flux from gamma-ray flaring blazars \citep{antares_blazars} and GRBs \citep{antgrb}, 
as well as a diffuse muon neutrino flux from extragalactic sources \citep{ant_diffuse}.  
These limits, however, do not yet strongly
constrain HEN emission and ultra-high-energy cosmic-ray production in
relativistic outflows \citep{huemmer:12,li:12,he:12}.
The \lo~Scientific Collaboration
and \vo~placed limits on the GW emission in GRBs
\citep{burstGrbS5,abadie:10inspiralgrb,abadie:s6vsr23grb} and
SGRs \citep{abadie:11sgr,abbott:09sgr,abbott:08sgr}. The exclusion
distances of these searches were, however, not sufficiently large to
expect a GW detection.

The above HEN and GW searches used timing and sky location information
from observations of events in the electromagnetic spectrum.  A
potentially large subset of GW and HEN sources may be intrinsically
electromagnetically faint, dust-obscured, or missed by telescopes, but sufficiently
luminous in GWs and HEN to be detected. Such sources may include, but
are not limited to, partially or completely choked GRBs with, perhaps,
only mildly relativistic jets
\citep{andobeacom:05,razzaque:05b,murase:06,wang:07}, relativistic
shock breakout in compact CCSN progenitor stars \citep{waxman:01b}, and
cosmic string cusps \citep{damour,siemens:06,berezinsky:11}.

Searches for HENs and GWs from such events have thus far relied on
blind (i.e., untriggered) all-sky searches. \cite{abbasi:12allsky}
performed an all-sky search for point sources of HENs in IceCube data
and \cite{antares_ps} carried out a similar study with \ant\ data. 
{\lo} and \vo~have carried out a number of all-sky searches for GWs. The
most recent and most sensitive such search for model-independent GW
bursts was published in \cite{abadie:12burstallsky}, whereas the most
recent all-sky search for binary inspiral-merger is
\cite{abadie:12cbcallsky}. All-sky model-dependent constraints on
cosmic string GW emission have been placed by
\cite{abbott:09strings}. The sensitivity of such blind all-sky
searches is limited by a much larger background compared to searches
based on timing and sky locations from electromagnetic observations.

A search for \emph{temporally and spatially coincident} HEN and GW
signals is a strong alternative to electromagnetically triggered or blind all-sky
analyses that search for GWs or HENs individually. Such a search is
independent of bias from electromagnetic observations, but still enjoys a much
reduced background thanks to timing and sky location constraints.  A
similar idea has been used in the follow-up of candidate GW events 
by the low-energy neutrino detector LVD \citep{LVD2004}.
A joint GW-HEN search was first proposed by \cite{asoY} and \cite{Prad}, 
and \cite{BartosObs} have derived constraints on joint GW-HEN signals 
based on the interpretation of independent GW and HEN observational results.
Here we present the first direct search for
coincident GW-HEN events, using data taken by the \ant~HEN
telescope and by the \lo~and \vo~GW observatories from January to
September 2007. At this time, \ant~was still under construction and
operating with only 5 active lines. At the same time, the fifth
\lo~science run (S5) and the first \vo~science run (VSR1) were
carried out. This was the first joint run of the \lo-\vo~network with
the detectors operating near their design sensitivities.

The basic principle of the analysis presented here is that of a
``triggered'' search: HEN candidates are identified in the \ant~data,
then the GW data around the time of the HEN event are analyzed for a
GW incident from the HEN estimated arrival direction.
This method has been applied previously in searches for GWs
associated with GRB triggers \citep{burstGrbS5,abadie:s6vsr23grb}.
It has been shown to have a distance reach up to a factor of 2 larger 
\citep{abadie:s6vsr23grb} than a blind all-sky search of 
the GW data, due to the reduced background. The expected rate of detections 
depends also on the beaming of the trigger signal, since the triggered 
search is only sensitive to the subset of sources oriented towards Earth.  
\citet{thesisWas} compares the analysis method used in this paper to 
the \lo-\vo~blind all-sky search \citep{abadie:12burstallsky} and 
predicts a detection rate for the triggered search of between $\sim$0.1 
and $\sim$6 times that of the blind search for beaming angles in the range 
5$^\circ$ -- 30$^\circ$.  
These numbers are broadly consistent with estimates for the special 
case of dedicated matched-filter searches for compact binary coalescence 
signals associated with short GRBs 
\citep[eg][]{2012arXiv1206.0703C,2012arXiv1209.3027K,2012arXiv1210.3095D,2012arXiv1210.6362N} 
after rescaling for a smaller distance improvement factor (typically 
$\sim$1.3, due to the better inherent background rejection of these 
specialised searches).
In either case, most of the GW events found by the triggered search 
will be new detections not found by the all-sky blind search, illustrating 
the value of the triggered search even when the relative detection 
rate is low \citep{thesisWas}.

We analyze a total of 158 HEN events detected at times when two or
more of the \lo-\vo~detectors were operating.  \ant~is sensitive to
HENs with energies greater than $\sim100$\,GeV \citep{ant_nu_osc}.  The \lo-\vo~analysis
targets model-independent burst GW signals with durations
$\lesssim$\,1\,s and frequencies in the 60\,Hz to 500\,Hz band.  The GW
search is extended in frequency up to 2000\,Hz only for a subset of the
HEN events, because the computational cost of such a search with the
current GW analysis pipeline is prohibitive.  

Statistical analyses of the HEN sample show no sign of associated
GW bursts.

This paper is organized as follows. In Section~\ref{sec:sources} we
discuss sources of coincident HEN and GW emission and expected
prospects for their detection. In Section~\ref{sec:detectors} we
describe the \ant, \lo, and \vo~detectors and the joint data taking
period.  Section~\ref{sec:HENselection} describes how the HEN sample
was selected.  Section~\ref{sec:GWselection} describes the search for
GWs coincident in time and direction with the HEN events.  We present
the results of the search in Section~\ref{sec:results}.  We discuss
the astrophysical implications of the results in
Section~\ref{sec:astrophysics} and conclude with considerations of the
potential for future joint GW-HEN searches.

\section{Candidate Sources for High-Energy Neutrino and 
Gravitational Wave Emission}
\label{sec:sources}

HEN emission is expected from baryon-loaded relativistic astrophysical
outflows. In the most common scenario (e.g., \citealt{waxman:97}),
Fermi-accelerated relativistic protons interact with high-energy
outflow photons in $p\gamma$ reactions leading to pions or kaons,
whose decay results in neutrinos, e.g., $\pi^+ \rightarrow \mu^+ +
\nu_\mu \rightarrow e^+ + \nu_e + \bar{\nu}_\mu + \nu_\mu$, which is
the dominant process (see, e.g., \citealt{winter:12}). This gives
$(\nu_e + \bar{\nu}_e:\nu_\mu + \bar{\nu}_\mu:\nu_\tau +
\bar{\nu}_\tau)$ production ratios of $(1:2:0)$, changing to
approximately $(1:1:1)$ at Earth due to flavor oscillations (e.g.,
\citealt{athar:06}). The HEN spectrum depends on the spectrum of the
accelerated protons (e.g.,
\citealt{guetta:04,abbasi:10grb,huemmer:12}) and, thus, on the
properties of the astrophysical source. In this section, we provide
estimates of the sensitivity of the 5-line \ant\ detector for HENs
from the various potential sources by estimating the probability $P =
X\%$ that at least one HEN is detected for a source at a given
distance $d_X$.

GW emission occurs, at lowest and generally dominant order, via
accelerated quadrupolar mass-energy dynamics. The coupling constant in
the standard quadrupole formula for GW emission (e.g.,
\citealt{thorne:87}) is $G c^{-4} \approx
10^{-49}\,\mathrm{s}^{2}\,\mathrm{g}^{-1}\,\mathrm{cm}^{-1}$ and the
directly detectable GW strain scales with (distance)$^{-1}$. For
example, a source at $10\,\mathrm{Mpc}$ needs a quadrupole moment of
$\sim 1\, M_\odot \times (100 \,\mathrm{km})^2$ that is changing on a
millisecond timescale to be detectable by a GW detector sensitive to a
strain of $10^{-21}$.  Equivalently, the minimum GW energy emission
detectable by the {\lo}-{\vo} network at this distance is
approximately $E_\mathrm{GW} \simeq 10^{-2}\,M_\odot c^2$ to $10\,M_\odot c^2$ for
frequencies between 100 Hz and 1000 Hz \citep{abadie:12burstallsky}.

In the following, we discuss a number of astrophysical scenarios in
which both HENs and GWs may be emitted at detectable levels.

\subsection{Canonical Long Gamma-Ray Bursts from Massive Stars}
\label{sec:introLGRBs}
\label{sec:lgrb}

Long-duration GRBs (LGRBs; $T_\mathrm{90} \gtrsim 2\,\mathrm{s}$;
$T_\mathrm{90}$ is the time over which $90\%$ of the $\gamma$ counts
are detected) are observationally implicated to be related to the
collapse of massive stars normally leading to core-collapse supernova
explosions \citep{hjorth:11,modjaz:11}. Typical LGRBs are strongly
beamed and most likely have jets with Lorentz factors $\Gamma \gtrsim
100$ and isotropic equivalent luminosities of 
$10^{51}\,\mathrm{erg}\,\mathrm{s}^{-1}$ to $10^{53}\,\mathrm{erg}\,\mathrm{s}^{-1}$
\citep{piran:05,meszaros:06,gehrels:09}. LGRBs are detected at a rate
of order $1$/(few days) by $\gamma$-ray monitors on satellite
observatories such as Swift/BAT \citep{gehrels:04,barthelmy:05} and
Fermi/GBM \citep{meegan:09gbm,bhat:04}.  It is important to note,
however, that Swift/BAT and Fermi/GBM miss $\sim$90\% and $\sim$40\%
of the prompt emission of all GRBs, respectively. This is due to
limited fields of view, technical downtime, and orbital passes through
the South Atlantic Anomaly. A nearby (say $\sim$10\,Mpc) LGRB will
have a bright multi-wavelength afterglow and may be accompanied by a
CCSN (see \S\ref{sec:llgrb}), but a significant fraction of local
CCSNe may have been missed by CCSN surveys based on galaxy catalogs
\citep{mattila:12}.

The central engine of LGRBs is expected to either be a
\emph{collapsar} (a black hole with an accretion disk;
\citealt{woosley:93,macfadyen:99}) or a \emph{millisecond magnetar}
(an extremely rapidly spinning, extremely magnetized neutron star;
e.g., \citealt{metzger:11}). In both scenarios, HEN emission may
result from a relativistic expanding fireball. HENs may begin
to be produced even before the jet breaks out of the stellar envelope
\citep{razzaque:03prd} and may continue well into the afterglow phase
\citep{murase:06b}.

HEN emission from canonical LGRBs is expected to have appreciable flux
for energies in the range $100\,\mathrm{GeV}$ to $100\,\mathrm{TeV}$. For
a LGRB at $\sim$$50\,\mathrm{Mpc}$ ($\sim$$10\,\mathrm{Mpc}$) one
would expect of order $1$ ($100$) HEN events in a
$\mathrm{km}^3$-scale water- or ice-Cherenkov detector (e.g.,
\citealt{waxman:97,guetta:04,huemmer:12,abbasi:10grb}).  
Based on the flux predictions of \cite{guetta:04}, the probability
for detection in the 5-line \ant~detector can be estimated to be
$\sim$50\% for a source at 10\,Mpc, which decreases to $\sim$$2\%$
for a source at $50\,\mathrm{Mpc}$. Note that these are most likely
optimistic estimates, since more detailed analyses suggest lower
HEN fluxes from GRBs \citep[e.g.,][]{huemmer:12}.

The most extreme scenario for GW emission in LGRBs is dynamical
fragmentation of a collapsing extremely rapidly differentially
spinning stellar core into a coalescing system of two protoneutron
stars \citep{davies:02,kobayashi:03}. Such a scenario may be unlikely
given model predictions for the rotational configuration of GRB
progenitor stars \citep[e.g.,][]{woosley:06}. Its GW emission,
however, would be very strong, leading to emitted energies
$E_\mathrm{GW} \sim 10^{-2}\, M_\odot c^2$ to $10^{-1}\,M_\odot c^2$ in the
$50\,\mathrm{Hz}$ to $1000\,\mathrm{Hz}$ frequency band of highest sensitivity of the
initial \lo/\vo~detectors, which could observe such an event out to
approximately $20\,\mathrm{Mpc}$ to $40\,\mathrm{Mpc}$ 
\citep{abadie:12cbcallsky,abadie:s6vsr23grb}.

In more moderate scenarios backed by computational models, GW emission
from LGRBs is likely to proceed, at least initially, in a very similar
fashion as in a rapidly spinning CCSN
\citep{ott:09,fryernew:11,kotake:12review}. If a black hole with an
accretion disk forms, a second phase of GW emission may come from
various hydrodynamic instabilities in the accretion disk
\citep[e.g.,][]{vanputten:04,Piro:2006ja,kiuchi:11}.

In the initial collapse of the progenitor star's core, a rapidly
rotating protoneutron star is formed. In this process, a
linearly-polarized GW signal is emitted with typical GW strains $|h|
\sim 10^{-21}$ to $10^{-20}$ at a source distance of $10\,\mathrm{kpc}$
and emitted energies $E_\mathrm{GW} \sim 10^{-8}\,M_\odot
c^2$ to $10^{-7}\,M_\odot
c^2$ between $100\,\mathrm{Hz}$ and $1000\,\mathrm{Hz}$ \citep{dimm,ott:12a}.
This part of the GW signal will only be detectable for Galactic events
and is thus not relevant here.  

In its early evolution, the protoneutron star (or protomagnetar,
depending on its magnetic field) may be spinning near breakup. This
can induce various rotational instabilities that induce ellipsoidal
deformations of the protoneutron star, leading to strong,
quasi-periodic, elliptically-polarized GW emission
\citep{fryer:02,corsi:09,ott:09,scheidegger:10b,fryernew:11}.  A
typical GW strain for a deformed protoneutron star of
$1.4\,M_\odot$ and radius of $12\,\mathrm{km}$, spinning with a period of
$1\,\mathrm{ms}$ may be $h \sim \mathrm{few} \times 10^{-22}$ at
$10\,\mathrm{Mpc}$. If the deformation lasted for $100\,\mathrm{ms}$,
$E_\mathrm{GW} \sim 10^{-1}\, M_\odot c^2$ would be emitted at
$2000\,\mathrm{Hz}$ \citep{fryer:02}.

In the collapsar scenario, accretion onto the protoneutron star
eventually leads to its collapse to a spinning black hole
\citep{oconnor:11}. This and the subsequent ringdown of the newborn
black hole leads to a GW burst at $\mathrm{few}\,\times10^2\,\mathrm{Hz}$ 
to $\mathrm{few}\,\times10^3\,\mathrm{Hz}$ with $h \sim 10^{-20}$ at
$10\,\mathrm{kpc}$ and $E_\mathrm{GW} \sim 10^{-7}\,M_\odot c^2$.  It
is thus detectable only for a Galactic source \citep{ott:11}.  

More interesting are hydrodynamic instabilities in the accretion
disk/torus that forms after seconds of hyperaccretion onto the newborn
black hole. The inner parts of the disk are hot, efficiently neutrino
cooled and thus thin \citep[e.g.,][]{popham:99} while the outer
regions are inefficiently cooled and form a thick accretion
torus. Gravitational instability may lead to fragmentation of this
torus into one or multiple overdense regions that may could condense
to neutron-star-like objects and then inspiral into the central black
hole \citep{Piro:2006ja}. For a source at $10\,\mathrm{Mpc}$, a $1
M_\odot$ fragment and a $8 M_\odot$ central black hole, this would
yield strains of $h \sim \mathrm{few} \times 10^{-22}$ and emitted
energies in the most sensitive band of $\sim 10^{-3}\,M_\odot c^2$ 
to $10^{-2}\,M_\odot c^2$.

The accretion torus may be unstable to the Papaloizou-Pringle
instability or to co-rotation-type instabilities
\citep{papaloizou:84,papaloizou:91}.  \cite{kiuchi:11} estimated $h
\sim 10^{-21}$ to $10^{-20}$ for a source at $10\,\mathrm{Mpc}$ and GW
frequencies of $100\,\mathrm{Hz}$ to $200\,\mathrm{Hz}$ for a $m = 1$--dominated 
non-axisymmetric disk 
instability in a disk around a $10\,M_\odot$ black hole. This
corresponds to $E_\mathrm{GW}$ of order $10^{-2}\,M_\odot c^2$ 
to $10^{-1}\,M_\odot c^2$.

In the speculative suspended accretion model for GRB accretion disks
\citep{vanputten:04}, low-order turbulence powered by black-hole
spindown may emit strong GWs. In the frequency domain,
this results in an anti-chirp behavior, since most of the emission is
expected to occur near the innermost stable orbit, which moves out in
radius as the black hole is spun down. Simple estimates suggest GW
strains $h \sim 10^{-21}$ at $10\,\mathrm{Mpc}$ and frequencies in the
$100\,\mathrm{Hz}$ to $1000\,\mathrm{Hz}$ band.  Depending on the initial black hole
spin, $E_\mathrm{GW}$ could be of order $1\, M_\odot c^2$.

\subsection{Low-Luminosity GRBs and Engine-Driven Supernovae}
\label{sec:llgrb}

Low-luminosity GRBs (LL-GRBs; also frequently referred to as X-ray
flashes) form a subclass of long GRBs with low $\gamma$-ray flux 
(e.g., \citealt{coward:05,hjorth:11,modjaz:11}). LL-GRBs are much more easily
missed by observations than LGRBs (see \S\ref{sec:lgrb}) and the small
observable volume (due to their low luminosity) suggests an event rate
that may be significantly higher than the rate of canonical
LGRBs \citep{soderberg:06,2007MNRAS.382L..21C,Le07,2007ApJ...662.1111L,Virgili07}. Five of the seven GRBs that have been unambigously associated
with a CCSN are LL-GRBs
\citep{hjorth:11,modjaz:11,zhang:12}. Moreover, all GRB-CCSNe are
highly energetic and of the spectroscopic type Ic-bl subclass. Ic
indicates a compact hydrogen/helium poor progenitor star and the
postfix -bl stands for ``broad line,'' because they have
relativistically Doppler-broadened spectral features.

Type Ic-bl CCSNe occur also without LL-GRB or LGRB, but are frequently
identified as \emph{engine-driven CCSNe} that exhibit luminous
radio emission \citep[e.g.,][]{soderberg:06,soderberg:10}.

Theory suggests (e.g.,
\citealt{khokhlov:99,burrows:07,lazzati:12,bromberg:11a}) that the
transition between engine-driven CCSNe, LL-GRBs with CCSNe, and
canonical LGRBs may be continuous. All are likely to be driven by a
central engine that launches a collimated bipolar jet-like outflow and
their variety may simply depend on the power output and duration of
central engine operation \citep{lazzati:12,bromberg:11a}. The power
output of the engine determines the energy of the jet and its Lorentz
factor. The duration of the central engine's operation determines if
the jet can leave the progenitor star and make a normal LGRB. If it
fails to emerge, the LGRB is ``choked'' and a more isotropic energetic
CCSN explosion is likely to result. As suggested by
\cite{bromberg:11a}, the relativistic shock breakout through the
stellar surface could then be responsible for a LL-GRB.

The GW emission processes that may be active in LL-GRBs and
engine-driven CCSNe are most likley very similar to the LGRB case
discussed in \S\ref{sec:introLGRBs} and we shall not consider them
further here. HEN emission is expected from the entire range of
stellar collapse outcomes involving relativistic flows. Since LL-GRBs
and engine-driven CCSNe are most likely much more frequent that
canonical LGRBs, much effort has been devoted to understanding the HEN
emission from such events \citep{waxman:01a,waxman:01b,razzaque:03prd,
  andobeacom:05,razzaque:05a,razzaque:05b,murase:06,murase:06b,wang:07,horiuchi:08}.
It is worthwhile to consider the probability of detection of HENs in
the 5-line \ant\ detector from LL-GRBs and engine-driven CCSNe, in
which mildly relativistic jets are likely to be involved.  The
detection probability depends strongly on the energy and the Lorentz
factor of the jet. Using the reference parameters of \cite{andobeacom:05}, 
$\Gamma = 3$
and $E_{\textrm{jet}} = E_0 \approx 3 \times 10^{51}$ erg, the detection
probability is $\sim$$50\%$ at $d_{50} =$ 1\,Mpc.

\subsection{Mergers and Short Gamma-Ray Bursts}

Short-duration GRBs (SGRBs; $T_\mathrm{90} \lesssim 2\,\mathrm{s}$)
are rarer than LGRBs and expected to result from double neutron star
(NS-NS) and/or neutron star -- black hole (BH-NS) mergers
\citep[e.g.,][]{nakar:07,gehrels:09}.  The isotropic equivalent energy
of SGRBs is 2 to 3 orders of magnitude smaller than the energy of
LGRBs.  Their jets have most likely lower Lorentz factors of $\Gamma
\sim 10$ to $50$ and wider opening angles.  Due to their short duration and
low isotropic equivalent energy, SGRBs are much easier to miss
observationally than LGRBs and their observable volume is much
smaller.

The efficiency of HEN emission in SGRBs depends on the efficiency of
proton acceleration, the $\gamma$-ray flux, and the SGRB variability
time scale \citep{nakar:07}. For a simple estimate of the detection
probability in the 5-line \ant~detector, one may resort to the HEN
flux estimates of \cite{guetta:04} (but see \citealt{huemmer:12} for
refined results). Assuming a jet with $\Gamma = 300$,
$E_\mathrm{jet} = 2 \times 10^{50}$ erg, one finds a distance
$d_{10} \sim 10\,\mathrm{Mpc}$ at which the probability of HEN
detection by the 5-line \ant\ detector is $10\%$. Hence, only the
closest and/or the most powerful SGRBs may be detectable.

The GW emission from NS-NS and BH-NS mergers is well studied
\citep[see][for reviews]{shibata:11,faber:12}. Most of the emission
comes from the late inspiral and merger phase during which the binary
sweeps through the $50\,\mathrm{Hz}$ to $1000\,\mathrm{Hz}$ band of 
highest sensitivity of \lo/\vo. The total emitted $E_\mathrm{GW}$ is 
of order $10^{-2}\,M_\odot c^2$ to $10^{-1}\,M_\odot c^2$.  
At the time of this analysis the \lo/\vo~network had maximum sensitive 
distances of $\sim$$30\,\mathrm{Mpc}$ for equal-mass NS-NS binaries and
$\sim$$50\,\mathrm{Mpc}$ for a BH-NS binary with a mass ratio of $4:1$, 
and a dedicated merger search on this data did not 
find any evidence for GW candidates \citep{s5vsr1cbc}.

\subsection{Bursting Magnetars}

Soft-gamma repeaters (SGRs) and anomalous X-ray pulsars are
X-ray pulsars with quiescent soft ($2-10\,\mathrm{keV}$) periodic
X-ray emissions with periods ranging from 5 to 10\,s. They exhibit
repetitive outbursts lasting $\sim$$0.1\,\mathrm{s}$ which reach peak
luminosities of $\sim$$10^{42}\,\mathrm{erg\,s}^{-1}$ in X-rays and
$\gamma$-rays. There are a number of known SGRs and anomalous X-ray pulsars 
\citep{hurley:11,mereghetti:11}, some of which have had rare hard
spectrum giant flares with luminosities of up to
$10^{47}\,\mathrm{erg\,s}^{-1}$.  The favoured model for these objects
is a magnetar, a neutron star with an extreme magnetic field of $B
\sim 10^{15}$\,G. Giant flares are believed to be caused either by
magnetic stresses fracturing the magnetar crust and leading to a
large-scale rearrangement of the internal field \citep{thompson:95} or
by a large-scale rearrangement of the magnetospheric field due to
magnetic reconnection \citep{lyutikov:06,gill:10}.  The sudden release
of energy and magnetic field rearrangement lead to the creation and
acceleration of pair plasma that may have some baryon loading, thus
leading to the emission of HENs in $p\gamma$ reactions
\citep{halzen:05}.  \cite{ioka:05} estimated the detectability of the
2004 giant flare of the Galactic SGR 1806-20 by HEN detectors. They
found that detectors such as IceCube and {\ant} should detect
multiple HEN events from similar Galactic SGR eruptions, provided the
baryon loading is sufficiently high. The AMANDA-II detector, which was
operating during the giant flare of SGR 1806-20, did not detect HENs
\citep{achterberg:06}. A search of IceCube data for HENs from regular
(non-giant) Galactic SGR flares also found no significant coincident
events \citep{abbasi:12allsky}. Estimates based on \cite{ioka:05} for
the 5-line \ant\ detector show that, $d_{50} \approx
200\,\mathrm{kpc}$ for baryon-rich flares, suggesting that similar
flares could be detected from anywhere within the Galaxy.

Significant emission of GWs in SGR giant flares may come from the
potential excitation of nonradial pulsational modes with
kHz-frequencies in the magnetar
\citep{freitas:98}. \cite{ioka:01} and \cite{corsi:11} placed
theoretical upper limits on the possible strength of the GW emission
based on the energy reservoir associated with a change in the
magnetic potential energy of the magnetar. They found an upper limit
for the emitted GW energy of $10^{-7}\,M_\odot c^2$ to $10^{-6}\,M_\odot c^2$, which
can be probed by the LIGO/Virgo network for a Galactic source
\citep{abbott:08sgr,abadie:11sgr}.  However, studies that investigated
the excitation of magnetar pulsational modes in more detail suggest
much weaker emission that may not be detectable even with
advanced-generation GW observatories \citep{levin:11sgr,zink:12}.

\subsection{Cosmic String Kinks and Cusps}

Cosmic strings are topological defects that may form in the early
Universe and are predicted by grand unified theories and superstring
theory \citep[e.g.,][]{kibble:76,polchinski:04}. Cosmic strings form
initially as smooth loops, but through interactions and
self-interactions may develop kinks and cusps
\citep[e.g.,][]{polchinski:04}. The kinks and cusps decay, which is
expected to lead to ultra-high energy cosmic ray emission with
energies in excess of $\sim 10^{11}\,\mathrm{GeV}$ and up to the
Planck scale \citep{hill:87,bhattacharjee:89}, including
ultra-high-energy neutrinos (UHENs; e.g.,
\citealt{anchordoqui:06,berezinsky:11,lunardini:12}) and GW
bursts \citep[e.g.,][]{damour,damour:01,cuesta:01,siemens:06}.

While not designed specifically for UHENs, HEN detectors like
\ant\ and IceCube have some sensitivity to UHENs in the $\gtrsim
10^{11}\,\mathrm{GeV}$ energy range. \cite{alvarez:01} predict,
depending on details of the underlying emission model, up to a few
events per year for a $\mathrm{km}^3$-scale detector. 
Since Earth is opaque to UHENs, downgoing events must be
selected. Since we are only considering {\ant} data for
neutrinos that have passed through Earth (see \S\ref{sec:antares}),
the present data set does not contain any potential UHEN events.

\cite{abbasi:11uhen}, who searched a year of IceCube-40 data for very
energetic HENs, did not find neutrinos in the $10^6\,\mathrm{GeV}$ to 
$10^9\,\mathrm{GeV}$ energy range, but did not report limits on UHENs.
A number of dedicated UHEN experiments exist, including ANITA
\citep{gorham:10anita}, NuMoon \citep{scholten:09} and others, but
have not yet constrained many emission scenarios from cosmic strings
\citep[e.g.,][]{lunardini:12}.

The rate of GW bursts from a network of cosmic strings depends on the 
string tension and other network parameters, and individual bursts may 
be detectable with advanced detectors \citep{damour:01,siemens:06}. 
The burst shape is expected to be generic, so that
matched-filtering GW analysis approaches may be employed. A first
search for GW bursts from cosmic string cusps in 15 days of LIGO data
from early 2005 did not reveal any candidate events
\citep{abbott:09strings}.

\section{GW and HEN detectors}
\label{sec:detectors}

\subsection{The \ant~neutrino telescope}
\label{sec:antares}
Since the Earth acts as a shield against all particles except neutrinos, 
a neutrino telescope mainly uses the detection of upgoing muons as a 
signature of muon-neutrino charged-current interactions in the matter around the
detector. The \ant~detector (Astronomy with a Neutrino Telescope and Abyss 
environmental RESearch) is currently the only deep sea high-energy-neutrino 
telescope and is operating in the Northern hemisphere \citep{Antares2}. 
The telescope covers an area of about 0.1\,km$^2$ on the sea bed, at a depth 
of 2475\,m, 40\,km off the coast of Toulon, France.  
The detector is a three-dimensional array of photomultiplier 
tubes (PMTs) \citep{ant_pmt}, hosted in pressure resistant glass spheres, called optical 
modules (OMs) \citep{ant_om}.  In its full configuration, it is composed of 12 detection 
lines, each comprising up to 25 triplets of PMTs, storeys, regularly 
distributed along 350\,m, the first storey being located 100\,m above the sea bed.
The first detection line was installed and connected in early 2006; the 
second line was put in operation in September 2006 and three more lines 
were connected in January 2007, so that a total of 5 lines were taking 
data in 2007. Five additional lines, together with an instrumentation line
(containing an ensemble of oceanographic sensors dedicated to the measurement 
of environmental parameters), were connected by the end of 2007. The telescope 
reached its nominal configuration, with 12 lines immersed and taking data, 
in May 2008.

The three-dimensional grid of PMTs is used to measure the 
arrival time and position of Cherenkov photons induced by the 
passage of relativistic charged particles through the sea water. 
This information, together with the characteristic emission angle of 
the light (about 43 degrees), is used to determine the direction of 
the muon and hence infer that of the incident neutrino.
The accuracy of the direction information allows to distinguish upgoing 
muons, produced by neutrinos, from the overwhelming background 
from downgoing muons, produced by cosmic ray interactions in the atmosphere 
above the detector \citep{ant_5lines}. Installing the detector at great depths serves to 
attenuate this background and also allows to operate the PMTs in a dark environment. At high energies the large muon range makes
the sensitive volume of the detector significantly greater than the 
instrumented volume. By searching for upgoing muons, the total \ant~sky coverage is 3.5$\pi$\,sr, with most of the 
Galactic plane being observable and the Galactic Center being visible 70\% 
of the sidereal day.

\subsection{Network of interferometers}

\subsubsection{LIGO}
\lo~is a network of interferometric gravitational wave detectors consisting of three
interferometers in the USA.  
These detectors are all kilometer-scale power-recycled Michelson laser interferometers with 
orthogonal Fabry-Perot arms \citep{Ligo1} able to detect the quadrupolar strain in space produced 
by the GW. Multiple reflections between mirrors located at the end points of each arm extend 
the effective optical length of each arm, and enhance the sensitivity of the instrument.

There are two \lo~observatories: one located at Hanford, WA and the 
other at Livingston, LA. The Hanford site houses two interferometers: 
one with 4\,km arms, denoted H1, and a second with 2\,km arms, denoted H2. 
The Livingston observatory has one 4\,km interferometer, L1. The 
observatories are separated by a distance of 3000\,km, corresponding to 
a time-of-flight separation of 10\,ms.

The \lo~instruments are designed to detect gravitational waves with 
frequencies ranging from $\sim40$\,Hz to several kHz, with a maximum 
sensitivity near 150\,Hz (see Fig \ref{fig:spectra}). In fact, seismic noise dominates at lower frequencies and 
the sensitivity at intermediate frequencies is determined mainly by thermal noise, with contributions from other
sources. Above $\sim 200$\,Hz, laser shot
noise corrected for the Fabry-Perot cavity response yields
an effective strain noise that rises linearly with frequency. 
The average sensitivities of the H1 and L1 detectors during the second year of the S5 run were about 20\% better than
the first-year averages, while the H2 detector had about the same average sensitivity in both years. 

\subsubsection{\vo}
\begin{figure}[h!]
\begin{center}
\includegraphics[width=0.45\textwidth]{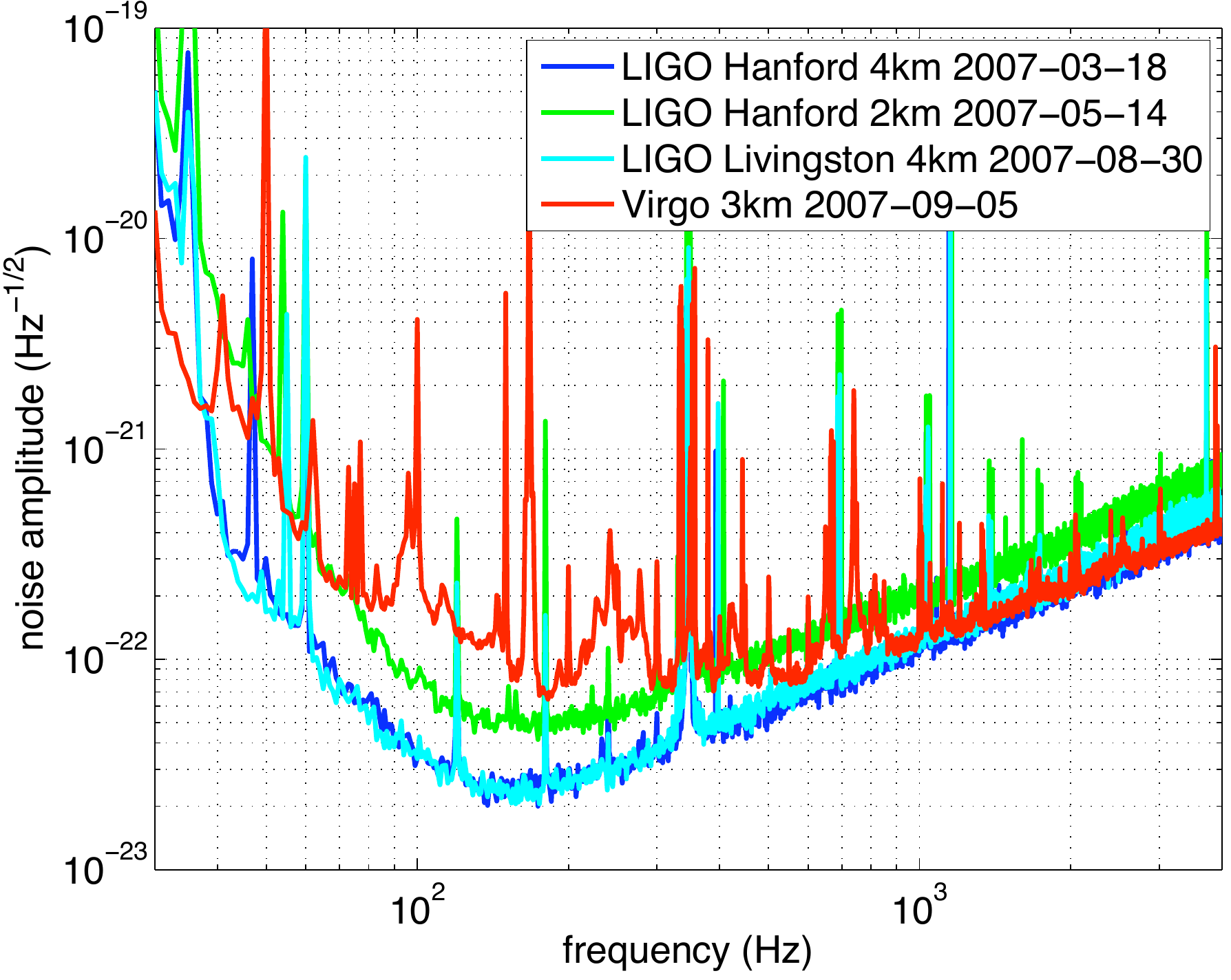}
\caption{Noise amplitude spectral densities of the four {\lo} and \vo~detectors during S5.}
\label{fig:spectra}
\end{center}
\end{figure}

The \vo~detector, V1, is in Cascina near Pisa, Italy. It is a 3\,km long 
power-recycled Michelson interferometer with orthogonal Fabry Perot arms 
\citep{virgo3}. 
The main instrumental difference with respect to \lo~is the seismic isolation system 
based on super-attenuators \citep{braccini}, chains of passive attenuators capable of
filtering seismic disturbances. The benefit from super-attenuators is a significant reduction 
of the detector noise at very low frequency ($<$40\,Hz) where \vo~surpasses the \lo~sensitivity.
During 2007, above 300\,Hz, the \vo~detector had sensitivity similar to the \lo~$4$\,km interferometers,
while above 500\,Hz it is dominated by shot noise, see Fig \ref{fig:spectra}.

The time-of-flight separation between the \vo~and Hanford observatories is 27\,ms, and 25\,ms between \vo~and Livingston.
Due to the different orientation of its arms, 
the angular sensitivity of \vo~is complementary to that of the \lo~detectors, \vo~therefore enhances the sky coverage of the network. 
Moreover, simultaneous observations of multiple detectors are crucial to reject environmental and instrumental effects.

At the time of writing the \lo~and \vo~interferometers are undergoing upgrades to ``advanced'' configurations with distance sensitivity improved by 
approximately a factor of 10 \citep{AdvancedITF}.  The advanced instruments will commence operations around 2015.

\subsection{Joint data taking periods}

The fifth \lo~science run, S5 \citep{Ligo3}, was held from 
2005 November 4 to 2007 October 1. Over one year of science-quality 
data were collected with 
all three \lo~interferometers in simultaneous operation at their design 
sensitivity, with duty factors of $75\%$, $76\%$, and $65\%$ for H1, H2, 
and L1. 
The \vo~detector started its first science run, VSR1 \citep{Virgo1}, 
on 2007 May 18. The \vo~duty factor over VSR1 was $78\%$. 
During this period, \ant~was operating in its 5 line configuration.
The concomitant set of \ant~5-line (5L), 
VSR1 and S5 data covers the period between January 27 and September 30, 
2007; these data are the subject of the analysis presented here.

\section{Selection of HEN candidates} 
\label{sec:HENselection}

\subsection{HEN data sample}

The \ant~data sample used in the analysis is composed of runs from 
2007 selected according to various quality criteria, based
mainly on environmental parameters (e.g. sea current, counting rates), 
configuration and behaviour of the detector during the given 
run (e.g. duration of the run, alignment of the detector). 
Two basic quantities are used to characterise the counting rate of 
a given OM: the baseline rate ($^{40}$K activity and bioluminescence) and the burst fraction (flashes 
of light emitted by marine organisms). 
The baseline rate represents the most probable counting
rate of a given OM computed from the rate distributions in each PMT 
over the whole run (typically a few hours). 
The burst fraction 
corresponds to the fraction of time during which 
the OM counting rates exceed by more than 20\% the estimated baseline. 
The data selected for this search are required to have a baseline rate 
below 120 kHz and a burst fraction lower than 40\%, with 80\% of all 
OMs being active. With these quality criteria, 
the active time is 103.4 days out of the 244.8 days of the 5-line period. 
Finally, when restricting the data to the concomitant period with 
\lo/\vo, the remaining equivalent time of observation is 
$T_{\textrm{obs}}=91$ days.

\subsection{Trigger levels}
\label{sec:triggers}

The \ant~trigger system is multi-level \citep{ant_daq}. The first level is applied 
{\it in situ}, while the remaining levels intervene after all data 
are sent to the shore station and before being written on disk. 
Trigger decisions are based on calculations done at three levels. 
The first trigger level, L0, is a simple threshold of about 
0.3 photo-electron (pe) equivalent charge applied to the analog 
signal of the PMT. The second level trigger, L1, is based on two 
coincident L0 hits in the same storey within 20\,ns or hits with 
large charge ($\ge$ 3\,pe or 10\,pe depending on the configuration). 
The L2 trigger requires the presence of at least five L1 hits in a 2.2\,$\mu$s time window (roughly
the maximum muon transit time across the detector) and that each pair of L1 hits 
are causally related according to the following condition: 
$\Delta t_{ij} \leq d_{ij} n/c + 20$\,ns.
Here $\Delta t_{ij}$ and $d_{ij}$ are the time difference and 
distance between hits $i$ and $j$, $c$ is the speed of light 
in vacuum and $n$ is the index of refraction.

\subsection{Reconstruction strategy}

Hits selected according to the criteria described in Section~\ref{sec:triggers} 
are then combined to reconstruct tracks using their arrival time 
and charge as measured by the corresponding OM. 
Muons are assumed to cross the detector at the speed of light along a straight 
line from which the induced Cherenkov light originates.
The time and charge information of the hits in the PMTs is used in a minimisation procedure 
to obtain the track parameters, namely, its direction $(\theta, \phi)$ and the position $(x_0, y_0, z_0)$
of one track point at a given time $t_0$.
The reconstruction algorithm used for this analysis is a fast and robust method \citep{BBfit} 
which was primarily designed to be 
used on-line.
\subsubsection{Description of the algorithm}
The algorithm is based on a $\chi^2$-minimisation approach. 
Its strict hit selection leads to a high purity up-down separation 
while keeping a good efficiency. 
The exact geometry of the detector is ignored: the detector lines 
are treated as straight and the 3 OMs of each storey are considered 
as a single OM centered on the line.
Thus, the hit's altitude corresponds to the optical modules altitude. 
All hits at the same floor in coincidence within 20\,ns are merged 
into one hit. The time of the merged hit is that of the earliest 
hit in the group and its charge is the sum of the charges.
The algorithm uses the L1 hits as a seed for the 
hit selection. It requests a coincidence of 2 L1 hits in two 
adjacent floors within 80\,ns or 160\,ns in two next-to-adjacent 
floors. 
The quality of the reconstruction is measured by a $\chi^2$-like 
variable with {\it NDF} degrees of freedom, based on the time differences between the hit times $t_i$ and 
the expected arrival time $t^{\gamma}_i$ of photons from the track or bright-point (see Section~\ref{sec:criteriahen}).
The quality function is then extended with a term that
accounts for measured hit charges $q_i$ and the calculated photon travel distances $d_i^{\gamma}$:
\begin{equation} \label{eqchi2}
\chi^2 = \frac{1}{\textit{NDF}} \sum^{N_{\textrm{hit}}}_{i=1} \left[ \frac{\Delta t_i^2}{\sigma_i^2} + \frac{Q(q_i)}{\bar{q}} \frac{D(d^{\gamma}_i)}{d_0}\right].
\end{equation}
In this expression, $\sigma_i$ is the timing error, set to 
10\,ns for charges larger than 2.5\,pe and to 20\,ns otherwise. 
$\Delta t_i = t^{\gamma}_i - t_i$ is the time residuals 
between the hit time $t_i$ and the expected arrival time of 
the photons $t^{\gamma}_i$ from the muon track. 
In the second term, $\bar{q}$ is the average hit charge calculated from all hits which have been selected for the fit and
$d_0 = 50\,\mathrm{m}$ is the typical distance at which the signal in one PMT from a Cherenkov light front is of the order of 1 pe.
The function $Q(q_i)$ accounts for the angular acceptance of the OMs, while $D(d^{\gamma}_i)$ penalises large amplitude hits originating 
from large distance tracks. 
A proper cut on the fit quality parameter allows the isolation of a high purity neutrino sample, 
which is crucial in the subsequent analysis.

\subsubsection{Azimuthal degeneracy of the reconstruction}
\label{sec:mirrorimage}

For a particle trajectory reconstructed from a Cherenkov cone giving hits 
on only two straight detector lines, there always exists an alternative trajectory having an 
identical $\chi^2$ value, but a different direction. The degenerate 
trajectory is the mirror image of the original track in the plane
formed by the two lines. 
As a consequence, each event reconstructed with only two lines will have two 
equiprobable arrival directions, which must be taken into account 
during the subsequent GW analysis.

\subsection{Criteria for HEN event selection}
\label{sec:criteriahen}
The initial sample of reconstructed events contains both upgoing 
neutrino induced muons and downgoing muons from cosmic ray interactions in the atmosphere. 
Some of the atmospheric muons are misreconstructed as upgoing and the selection cuts, based on Monte-Carlo 
simulations, are devised to reduce this contamination so as to maximise the discovery potential.
A minimum of 6 hits on at least 2 lines are required to reconstruct a track.
Only upgoing tracks are kept for further analysis. 
Quality cuts are then applied based on two quantities computed 
according to equation~(\ref{eqchi2}). The first parameter used, 
$\chi^2_t$, is the quality factor associated with the reconstructed 
particle track, whereas the second one, $\chi^2_b$, is associated 
with a bright-point, light emitted from a point-like source inside the detector.
This rejects events from large electromagnetic showers, likely to appear in downgoing muon 
bundles for instance.

A cut on $\chi^2_b$ reduces the number of such events and decreases the contribution 
of misreconstructed muons in the background. Further cuts are applied 
on $\chi^2_t$ depending on the arrival direction of the candidate - 
the muon contamination increases close to the horizon - which 
reduce the fraction of misreconstructed muons to less than 20\% over 
the whole sample, while optimising the sensitivity (see Section \ref{sec:henanalysis} and \cite{GaroPhD}).

\begin{figure}[h!]
\begin{center}
\includegraphics[width=0.5\textwidth]{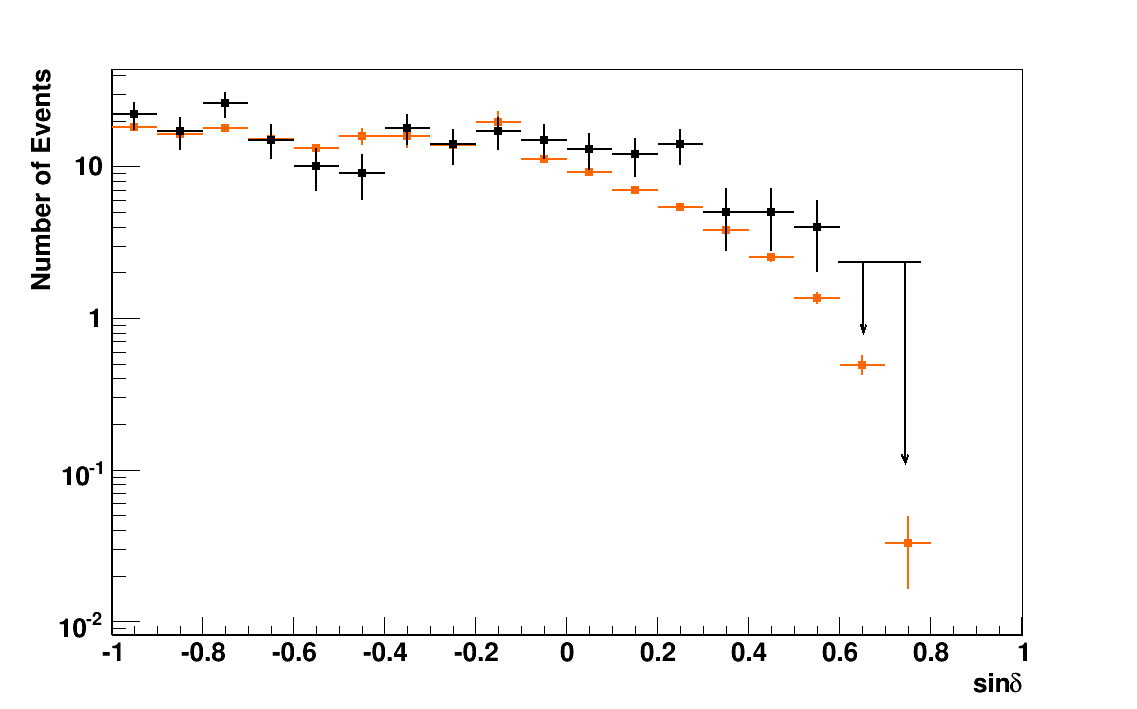}
\caption{Distribution of the sine of the declination $\delta$ of selected events (black points), 
compared to Monte-Carlo expectations (sum of atmospheric muons and atmospheric neutrinos, orange (or grey) points).
}
\label{fig:henselectedevents}
\end{center}
\end{figure}

Figure \ref{fig:henselectedevents} shows the distribution of the sine of the declination
of the events selected with the final cuts, which is globally consistent with background.

\subsection{Angular error}
\label{sec:angularerror}

\begin{figure}[h!]
\begin{center}
\includegraphics[width=0.5\textwidth]{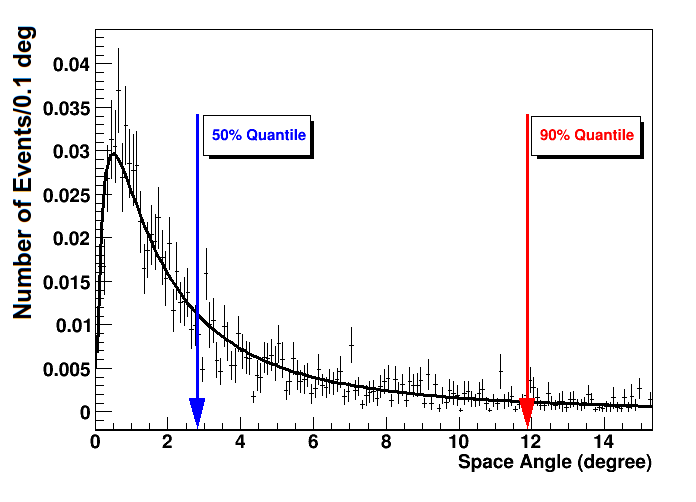}
\caption{Example of space angle distribution with the associated fit 
to equation (\ref{eq:lognormal}) obtained with a sample of Monte-Carlo HEN events, for a given declination and a given number of hits.  
The arrows indicate the 50$^\mathrm{th}$ (median) and the 90$^\mathrm{th}$ percentiles of the distribution. 
The distribution is normalised to unity.}
\label{fig:henspaceangle}
\end{center}
\end{figure}

The distribution of the space angle $\Omega$ between the true neutrino 
arrival direction and the reconstructed muon track can be described by a log-normal distribution:
\begin{equation}\label{eq:lognormal}
P(\Omega) = \frac{1}{\sqrt{2\pi}} \frac{e^{-\frac{1}{2 \sigma_0^2}\left(\ln \left( \frac{\Omega-\theta_0}{m_0}\right)\right)^2}}{\left(\Omega-\theta_0 \right)\sigma_0} \, , 
\end{equation}
where $\theta_0$ is a location parameter, $\sigma_0$ is related to the shape of the distribution and 
$m_0$ is a scaling parameter. 
In all cases for our study, the location parameter $\theta_{0}$ is close to zero, and
$(\Omega-\theta_{0})\;>0$ is always satisfied.
This distribution depends on the energy associated to the track 
(estimated through the number of photons detected) and its declination. 
This parametrisation is used during the GW search 
to compute the significance of a hypothetical signal for the 
scanned directions inside the angular search window centred 
around the reconstructed neutrino arrival direction. 
Figure \ref{fig:henspaceangle} shows an example of distribution of the 
space angle for a sample of Monte Carlo neutrinos with an $E^{-2}$ spectrum, together with 
the best-fit parametrisation and the 50$^\mathrm{th}$ and
90$^\mathrm{th}$ percentiles of the distribution.

One of the main variables to describe the performance of a neutrino telescope is the angular resolution, defined as the median of the 
distribution of the angle between the true neutrino direction and 
the reconstructed track, also indicated in Figure \ref{fig:henspaceangle}.  This number is estimated from simulations. 

For those events of our selected sample reconstructed with at least three lines the angular resolution is, 
assuming an $E^{-2}$ energy spectrum, $\sim 2.5^{\circ}$ at 100\,GeV, improving to $1^{\circ}$ around 100\,TeV.
For 2-line events, when 
selecting the reconstructed track closer to the true direction, the angular 
accuracy varies between 3$^{\circ}$ at low energy (100\,GeV) and $2.5^{\circ}$ at
high energy (100\,TeV). 

We define the angular search window for the GW analysis
as the 90$^\mathrm{th}$ percentile of the distribution, also indicated in Figure \ref{fig:henspaceangle}; this window 
lies between 5$^{\circ}$ and 10$^{\circ}$ for 3-line events, depending on declination, 
and between 10$^{\circ}$ and 15$^{\circ}$ for 2-line events. 

We note that the typical angular distance between galaxies within 10\,Mpc is 
a few degrees \citep{GWGC}, much smaller than the typical 
size of the 90$^\mathrm{th}$ percentile error region for our HEN events. 
This implies that we
can associate a potential host galaxy to any of the HEN candidates if it turns
out to be of cosmic origin.

\subsection{Analysis sensitivity and selected HEN candidates}
\label{sec:henanalysis}

The limit-setting potential of the analysis, or sensitivity, 
has been 
quantified for the whole 5 line data period. Specifically, the sensitivity 
is defined as the median 90\% upper limit obtained over an ensemble of 
simulated experiments with no true signal. 
The sensitivity depends on the declination of the potential source.
For our sample and 
assuming an $E^{-2}$ steady flux, using the selection criteria described, 
the best sensitivity has been estimated to be $E^{2} \frac{dN}{d E} \approx 10^{-6}\,\textrm{GeV}$\,cm$^{-2}$\,s$^{-1}$.  
This best sensitivity is reached below $-47^{\circ}$; i.e., at declinations which are 
always below the horizon at the latitude of {\ant} ($43^{\circ}$N).

\begin{figure*}[htb]
\includegraphics[width=\textwidth]{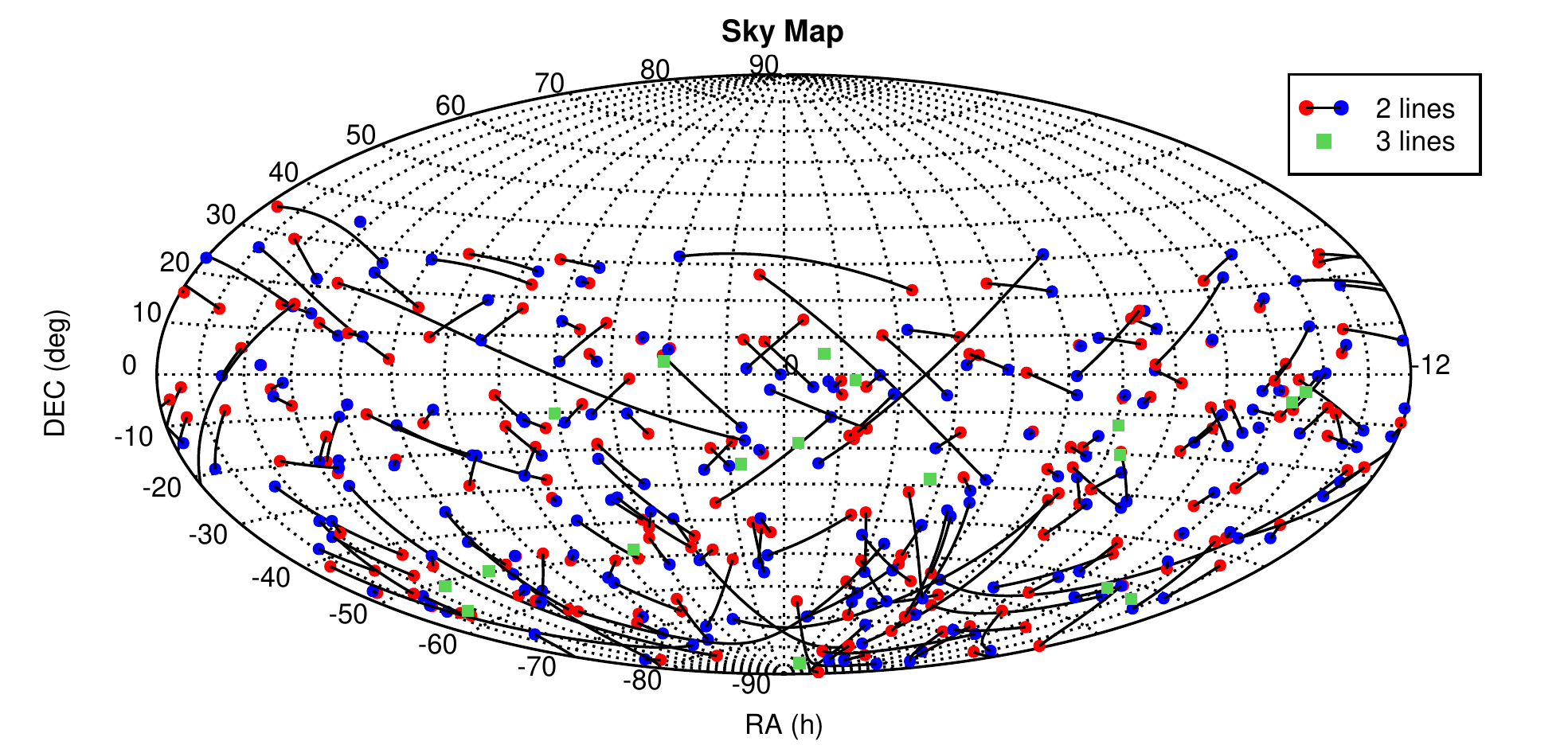}
\caption{Skymap of the selected 216 HEN events in equatorial coordinates. 
A line connects the associated mirror solutions for events reconstructed 
with two lines as described in Section~\ref{sec:mirrorimage}.}
\label{fig:maphen}
\end{figure*}

With the selection previously described, 181 runs corresponding to 104 days 
of live time were kept for the analysis.
The selection has been divided into 
events reconstructed with 2 lines and events with at least 3 lines. 
Each of the mirror solutions for 2 line events will be searched for 
possible counterparts in the subsequent GW analysis. This results 
in 216 neutrinos to be analysed: 198 with two
possible directions and 18 reconstructed 
with at least 3 lines. 
Figure~\ref{fig:maphen} is a sky map of the candidate HEN events, where 
the degenerate solutions for 2 line events can be seen.

Of these HEN events, 158 occurred at times when at least two 
gravitational-wave detectors were operating.  Since two or more 
detectors are required to discriminate GW signals from background 
noise (as described in Section~\ref{sec:GWevents}), in the following 
we consider only these remaining $158$ HEN candidates: $144$ 2-line 
events and $14$ 3-line events\footnote{Details of each of the HEN 
candidate events are given at 
\url{https://dcc.ligo.org/cgi-bin/DocDB/ShowDocument?docid=p1200006}.}.

Finally, we note that IceCube operated in its 22-string configuration 
for part of 2007 \citep{abbasi:12allsky}.  However, this data was only used for 
time-dependent searches applied to source directions with observed X-ray 
or gamma-ray emission, such as GRBs; there were no untriggered, 
time-dependent searches over the sky.  
Furthermore, a comparison of {\ant} and IceCube sensitivities in 2007 indicates 
that the bulk of our HEN neutrino triggers come from declinations 
(the southern sky) 
such that it is unlikely that IceCube could have detected the source 
independently.

\section{GW search method} 
\label{sec:GWselection}

\subsection{Search procedure}

One of the simplest searches that may be performed combining GW and 
HEN data is a triggered analysis that scans GW data around 
the time of the putative neutrino event by cross-correlating data 
from pairs of detectors. This search exploits knowledge of the time 
and direction of the neutrino event to improve the GW search sensitivity. 
We use the \xp~algorithm \citep{Xpipeline}, which has been used in 
similar searches for GWs associated with GRBs \citep{burstGrbS5,abadie:s6vsr23grb}. 
\xp~performs a coherent analysis of data from arbitrary networks of 
gravitational wave detectors, while being robust against background 
noise fluctuations. Each trigger is analysed independently of the others, 
with the analysis parameters optimised based on background noise 
characteristics and detector performance at the time of that trigger, 
thereby maximising the search sensitivity.

\subsection{GW event analysis}
\label{sec:GWevents}

In our GW search, a neutrino 
candidate event is characterised by its arrival time, direction, and angular
search window (and mirror-image window, for the 2-line events).
Also important is the range of possible time delays (both positive and 
negative) between the neutrino emission and the associated
gravitational-wave emission. This quantity is referred to as the 
{\it on-source} window for the neutrino; it is the time interval 
which is searched for GW signals.  We use a symmetric on-source 
window of $\pm$496\,s \citep{Gwhen1}, which is conservative enough 
to encompass most theoretical models of GW and HEN emission.
The maximum expected time delay between GWs and HENs
due to a non-zero mass effect for either particle is much smaller
than the coincidence windows used.

The basic search procedure follows that used in \cite{burstGrbS5}. 
All detectors operating at the time of the trigger and which pass 
data-quality requirements are used for the GW search. 
The data from each detector are first whitened and time-delayed 
according to the sky location being analysed so that a GW signal 
from that direction would appear simultaneous in each data stream.
The data are then Fourier transformed to produce time-frequency maps.  
The maps are summed coherently (using amplitude 
and phase) with weighting determined by each detector's 
frequency-dependent sensitivity and response to the sky location in question; 
the weightings are chosen to maximise the signal-to-noise ratio expected 
for a circularly polarized GW signal\footnote{
Empirically it is found that the circular polarisation restriction 
also improves the overall detection probability for {\em linearly} 
polarised GWs, as the resulting background reduction outweighs the 
impact of rejecting some linearly polarised GWs.  
},
which is the expected polarisation for a GW source observed from near 
the rotational axis \citep{kobayashi-2003-585}.
A threshold is placed on the map to retain the largest 1\% of pixels 
by energy (squared amplitude).  Surviving pixels are grouped using 
next-nearest-neighbours clustering; each cluster of pixels is 
considered as a candidate GW event. 
The event cluster is assigned a combined energy by summing the energy values 
of its constituent pixels; this combined energy is used as the ranking 
statistic for the events.

In addition to the marginalised circular polarization sum, 
a second ranking statistic is computed based on a 
maximum-likelihood analysis of the event assuming power-law 
distributed background noise with no assumption on the GW 
polarization. In practice this statistic is often found to provide 
signal-noise separation due to the non-Gaussian nature of the 
GW detector noise.
Other combinations of the data are also constructed.
Of particular importance are ``null'' combinations designed to 
cancel out the GW signal from the given sky location; comparison 
to corresponding ``incoherent'' combinations provides powerful  
tests for identifying events due to background noise fluctuations 
\citep{Ch_etal:06}, and are described in detail in \citet{s6methods}.
Events are also characterised by their duration, 
central time, bandwidth, and central frequency.

The time-frequency analysis is repeated for Fourier transform lengths 
of 1/128, 1/64, 1/32, 1/16, 1/8, 1/4\,s, to maximise the sensitivity 
to GW signals of different durations.  It is also repeated over a grid 
of sky positions covering the 90\% containment 
region of the HEN.  This grid is designed such that the maximum relative 
timing error between any pair of GW detectors is less that 0.5\,ms. 
When GW events from different Fourier transforms lengths or sky positions 
overlap in time-frequency, the highest-ranked event is kept and the others 
discarded.  Finally, the events are decimated to a rate of 
0.25\,Hz before being written to disk.

This time-frequency analysis is performed for all of the data in the 
$\pm$\,496\,s on-source window.  To estimate the significance of the resulting 
GW candidates, the same analysis is repeated for all coincident data in 
the {\it off-source} window, defined as all data within $\pm\,1.5$\,hours of the neutrino time, excluding the on-source interval. 
The same set of detectors and data-quality requirements as in the 
on-source analysis are used for the off-source data.
These off-source data provide a sample of background that 
does not contain any signal associated with the neutrino event, 
but with statistical features similar to the data searched in 
association with the neutrino.  To enlarge the background sample, 
we also repeat the off-source analysis after applying time shifts 
of multiples of 6 s to the data from one or more 
detectors; with such time slides we were able to produce $O(10^3)$ 
background trials for each HEN.

Finally, the analysis is repeated after ``injecting'' (adding) 
simulated GW signals to the on-source data.  The amplitudes and morphologies 
tested are discussed in Section~\ref{sec:injections}.  We use 
these simulations to optimise and assess the sensitivity of the search, 
as discussed below.

\subsection{GW search optimisation}
\label{sec:GWtuning}

The sensitivity of searches for gravitational-wave bursts tends to be 
limited by the presence of non-Gaussian fluctuations of the background 
noise, known as glitches.  To reduce this background, events that overlap 
in time within known instrumental and/or environmental disturbances 
are discarded.  In addition to this ``veto'' step, GW consistency 
tests comparing the coherent and incoherent energies are applied to 
each event \citep{s6methods}.  These tests are applied to the on-source, 
off-source and injection events; events failing one or more of these 
tests are discarded.  
The thresholds are optimised by testing a preset range of thresholds and 
selecting those which give the best overall detection efficiency 
at a fixed false alarm probability of 1\% when applied to a random 
sample of background and injection events (the on-source events are 
{\it not} used; i.e., this is a blind analysis).  These tests also 
determine which of the two ranking statistics discussed in 
Section~\ref{sec:GWevents} (based on circularly polarized GW energy or 
powerlaw noise) gives the better detection efficiency; the winner is 
selected as the final ranking statistic.

Once the thresholds have been fixed, these consistency tests are applied 
to the on-source events and to the remaining off-source and injection 
events (those not used for tuning).
The surviving on-source event with the largest significance 
(highest energy or powerlaw statistic) is taken to be the best candidate 
for a gravitational wave signal and is referred to as the loudest event 
\citep{badri}.
All surviving on-source events are assigned a false alarm probability 
by comparison to the distribution of loudest events from the off-source 
trials.  Any on-source event with probability $p<0.01$ is subjected to 
additional checks to try to determine the origin of the event 
and additional background time slide trials are performed to improve 
the accuracy of the false alarm probability estimate.

After the $p$ values have been determined for the loudest events associated 
with each of the 158 HEN events, the collective set of $p$ values is tested 
for consistency with the null hypothesis (no GW signal) using the binomial test, 
discussed in Section~\ref{sec:binomial}.  We also set a frequentist upper limit 
on the strength of gravitational waves associated with each neutrino trigger, 
as discussed in Section~\ref{sec:gwuls}.

\subsection{Low-frequency and high-frequency GW analyses}
\label{sec:lowvshigh}

Given our knowledge of possible GW sources discussed in 
Section~\ref{sec:sources}, the most likely detectable signals 
at extra-galactic distances are in the low-frequency band 
($f\lesssim500$\,Hz), where our detectors have maximum sensitivity, see Fig \ref{fig:spectra}.
At the same time, the computational cost of the \xp~analysis increases 
at high frequencies. This is due in part to the extra data to be 
analysed, but also to the need for finer-resolution sky grids to keep 
time delay errors much smaller than one GW period.  We therefore split 
the gravitational wave band into two regions: 60\,Hz to 500\,Hz and 500\,Hz to 2000\,Hz.  
The low-frequency band is analysed for all HEN events -- such a 
search is computationally feasible while covering the 
highest-sensitivity region of the GW detectors.  However, compact 
objects such as neutron stars or collapsar cores have characteristic 
frequencies for GW emission above 500\,Hz.  Such emissions might be 
detectable from Galactic sources such as soft gamma repeater giant 
flares, or possibly from nearby galaxies. 
Since the computational cost of a high-frequency search for all HEN 
events is prohibitive with the current analysis pipeline, 
we perform the 500\,Hz to 2000\,Hz analysis on the 3-line HEN events only.  
The 3-line events are a small subset ($\sim$10\%) of the total trigger 
list and have the smallest sky position uncertainties, and therefore 
the smallest computational cost for processing. 
To reduce the computational cost further, we use the same sky grid 
for the high-frequency search as was used at low frequencies, after 
determining that the loss of sensitivity is acceptable.
The high-frequency analysis is performed independently 
of the low-frequency analysis (independent tuning, background 
estimation, etc.) using the identical automated procedure. 
In the following sections we will present the results of the low-
frequency and high-frequency searches separately.

\section{Coincident search results}
\label{sec:results}

\subsection{Per-HEN GW candidates}

We analysed GW data in coincidence with 158 neutrino candidates 
for the low frequency search, and 14 neutrino events for the high 
frequency search. In the low frequency analysis, only one neutrino trigger 
had a corresponding GW event with false alarm probability below the 
threshold of $p=0.01$ to become a candidate event. We found no candidates 
in the high frequency search.
For the low-frequency candidate, additional time shifts totaling 18064 
background trials yielded a refined false alarm probability of $p=0.004$, 
which is not significant given a trials factor of 158 (this statement is quantified below). 
This event came from analysis of the H1, H2, and V1 
data; follow-up checks were performed, including checks of detector 
performance at the time as indicated by monitoring programs and operator 
logs, and scans of data from detector and environmental monitoring 
equipment to look for anomalous behaviour. 
While these checks did not uncover a physical cause for the event, 
they did reveal that it occurred during a glitching period in V1. 
We conclude that we have no clear gravitational wave burst signal 
associated with any of our sample of 158 neutrino events.

\subsection{Search for a cumulative excess: binomial test}
\label{sec:binomial}

A quantitative analysis of the significance of any candidate 
gravitational-wave event must take account of the trials factor due to the 
number of neutrino events analysed.  We use the binomial test, which has 
been applied in previous GRB-triggered GW searches 
\citep{multigrb07,burstGrbS5}.  
Under the null hypothesis, the false alarm probabilities $p$ for each HEN 
loudest event are expected to be uniformly distributed between 0 and 1. 
The binomial test compares the measured $p$ values to the null distribution 
to determine if there is a statistically significant excess of (one or more) small $p$ 
values which may be due to gravitational wave signals. 

Briefly, the binomial test sorts the set of $N$ measured loudest event 
probabilities in ascending order: $p_1 \le p_2 \le p_3 \le ... \le p_N$. 
For each $i\in[1,N_\textrm{tail}]$ we compute the binomial 
probability $P_{\ge i}(p_i)$ of getting $i$ or more events with $p$ 
values $\le p_i$: 
\begin{equation}
P_{\ge i}(p_i) = \sum_{k=i}^{N} \frac{N!}{(N-k)!k!}p_i^{k}(1-p_i)^{N-k} \, .
\end{equation}
Here $N$ is the number of HEN events analysed (158 in the 60\,Hz to 500\,Hz band and 14  
in the 500\,Hz to 2000\,Hz band), and $N_\textrm{tail}$ is the number of the smallest 
$p$ values we wish to test.  We choose $N_\textrm{tail}$ to be 5\% of $N$; 
i.e., $N_\textrm{tail} = 8$ for the low frequency band and 
$N_\textrm{tail} = 1$ for the high frequency band.

The lowest $P_{\ge i}(p_i)$ for $i\in[1,N_\textrm{tail}]$ is taken as the 
most significant deviation from the null hypothesis.  To assess the 
significance of the deviation, we repeat the test using $p$ values drawn 
from a uniform distribution and count the fraction of such trials which 
give a lowest $P_{\ge i}(p_i)$ smaller than that computed from the true 
measured $p$ values. 

\begin{figure}[h!]
\begin{center}
\includegraphics[width=0.48\textwidth]{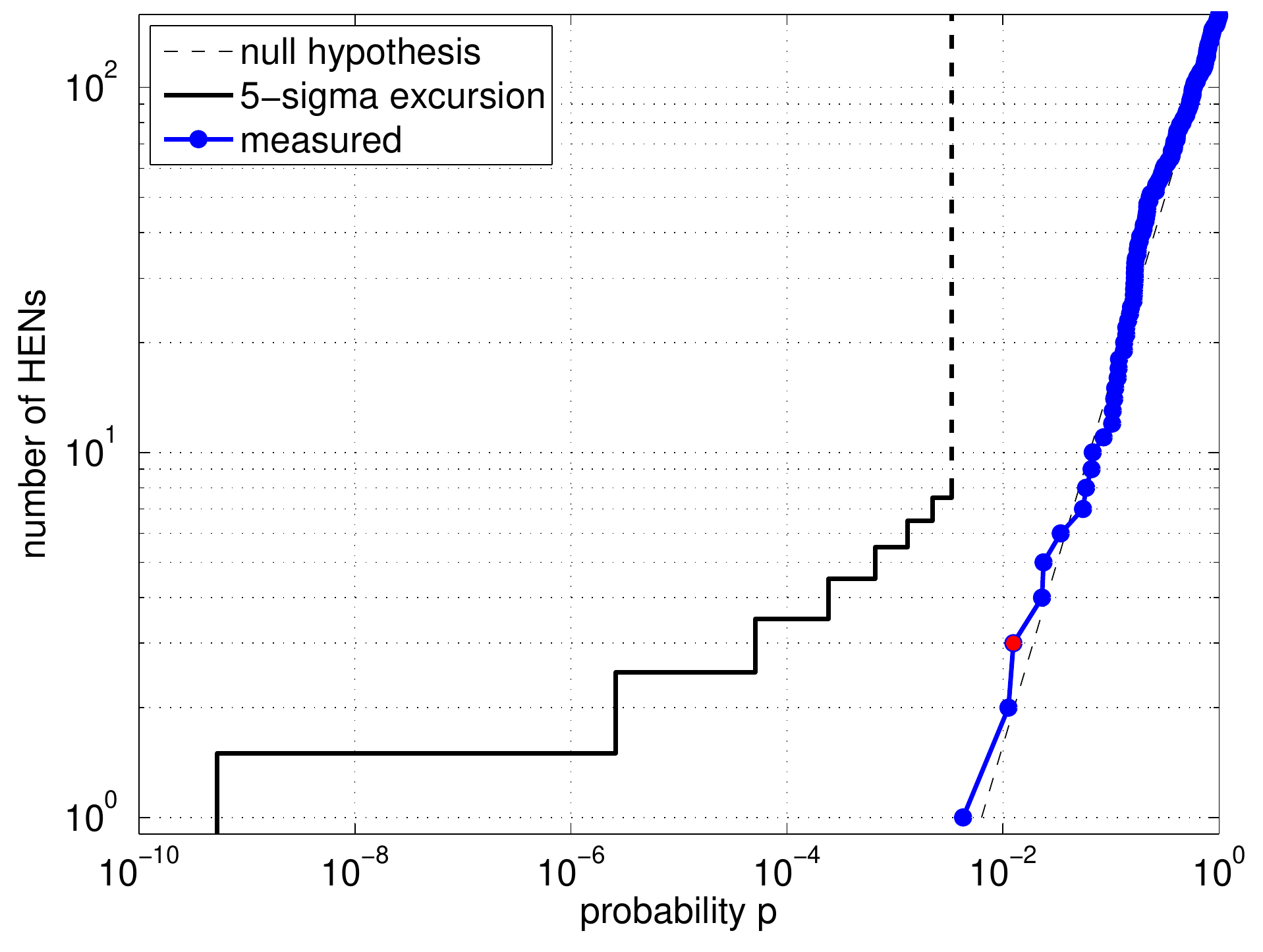}
\caption{Distribution of observed $p$ values for the loudest GW event 
associated with each neutrino analysed in the low frequency analysis. 
The red dot indicates the largest deviation of the low $p$ tail from 
the uniform distribution null hypothesis; this occurs due to having the 
three loudest events below $p_3\sim 0.013$.  Deviations this large or 
larger occur in approximately 64\% of experiments under the null hypothesis.
The black line shows the threshold for a 5-sigma deviation from the null hypothesis. 
}
\label{fig:bin_LF}
\end{center}
\end{figure}

\begin{figure}[h!]
\begin{center}
\includegraphics[width=0.48\textwidth]{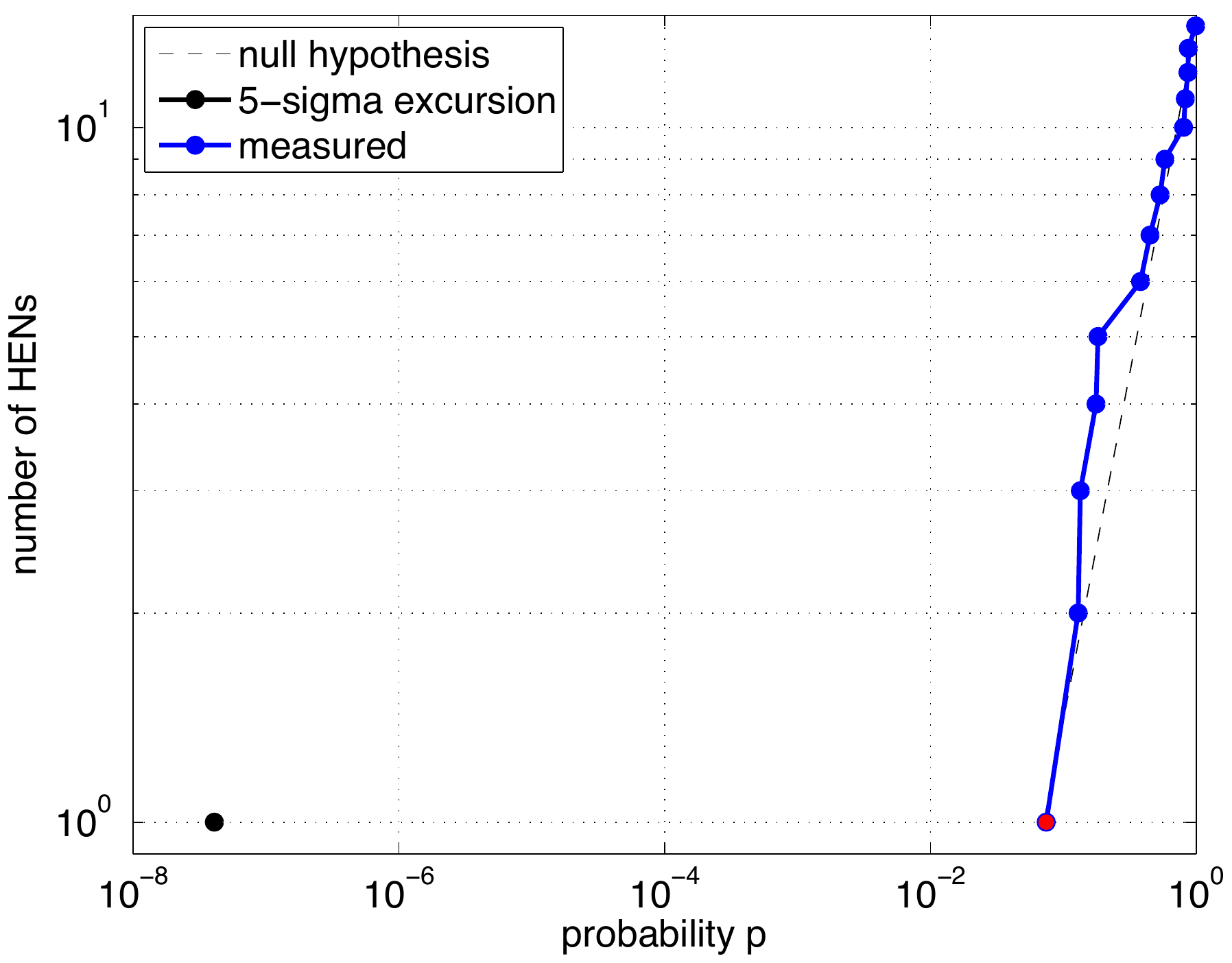}
\caption{Distribution of observed $p$ values for the loudest GW event 
associated with each neutrino analysed in the high frequency analysis. 
The red dot indicates the largest deviation of the low $p$ tail from 
the uniform distribution null hypothesis; since $N_\textrm{tail}=1$, 
this is constrained to occur for $p_1$.  Deviations this large or 
larger occur in approximately 66\% of experiments under the null hypothesis.
The black dot shows the threshold for a 5-sigma deviation from the null hypothesis. 
}
\label{fig:bin_HF}
\end{center}
\end{figure}

Figures~\ref{fig:bin_LF} and \ref{fig:bin_HF} show the cumulative 
distribution of $p$ values measured in the low- and high-frequency 
analyses.  
In both cases the measured $p$ values are consistent with the null hypothesis. 

\subsection{GW upper limits}
\label{sec:gwuls}

The sensitivity of the GW search is determined by a Monte-Carlo analysis. 
For each neutrino trigger, we add simulated GW signals to the on-source 
data and repeat the analysis described in Section~\ref{sec:GWevents}. 
We consider a simulated signal detected if it produces an event louder 
than the loudest on-source event after all event tests have been applied.  
We define a $90\%$ confidence level lower limit on the distance to the source as the maximum distance $D_{90\%}$ such that for any distance 
$D \le D_{90\%}$ the probability of detection is 0.9 or greater.

\subsubsection{Injected waveforms}
\label{sec:injections}

As in GRB-triggered searches, we use a mix of {\it ad hoc} and 
astrophysically motivated GW waveforms.  The {\it ad hoc} waveforms are 
Gaussian-modulated sinusoids:
\begin{eqnarray}
h_+ 
  & = & \frac{(1+\cos^2\iota)}{2}
        \frac{h_\textrm{rss}}{(2\pi \tau^2)^{\frac{1}{4}}} 
        \textrm{e}^{-\frac{(t-t_0)^2}{4\tau^2}} \cos{2\pi f_0(t-t_0)}
        \, , \quad \\
h_\times
  & = & \cos\iota  
        \frac{h_\textrm{rss}}{(2\pi \tau^2)^{\frac{1}{4}}} 
        \textrm{e}^{-\frac{(t-t_0)^2}{4\tau^2}} \sin{2\pi f_0(t-t_0)} \, .
\end{eqnarray}
Here $f_0$ is the central frequency, $t_0$ is the central time, and 
$\tau$ is the duration parameter.
This waveform is consistent with the GW emission from a rotating system 
viewed from an inclination angle $\iota$ to the rotational axis.  
We select the inclination uniformly in $\cos\iota$ with 
$\iota\in[0^{\circ},5^{\circ}]$.  This corresponds to a nearly on-axis 
system, such as would be expected for association with an observed long 
GRB.
We chose $\tau=1/f_0$, and use central frequencies of 
100\,Hz, 150\,Hz, and 300\,Hz for the low-frequency analysis and 554\,Hz and 
1000\,Hz for the high-frequency search. The quantity $h_\textrm{rss}$ is the 
root-sum-square signal amplitude:
\begin{equation}
h_\textrm{rss} \equiv \sqrt{\int\left(h_+^2(t)+h^2_{\times}(t)\right)dt} \, .
\end{equation}
For the small values of $\iota$ considered here ($\iota<5^{\circ}$) this 
amplitude is related to the total energy $E_\textrm{GW}$ in a narrow-band  
gravitational-wave burst by 
\begin{equation}
E_\textrm{GW} \simeq \frac{2}{5} \frac{\pi^2 c^3}{G} h_\textrm{rss}^2 f_0^2 D^2 \, .
\end{equation}

For astrophysical injections we use the gravitational-wave emission of  
inspiraling neutron star and black hole binaries, which are widely thought 
to be the progenitors of short GRBs. Specifically, we use the 
post-Newtonian model for the inspiral of a double neutron star system with component masses $m_1=m_2=1.35M_\odot$, and the one for a black-hole - 
neutron-star system with $m_1=5M_\odot$, $m_2=1.35M_\odot$. 
We set the component spins to zero in 
each case. Motivated by estimates of the jet opening angle for short GRBs, 
we select the inclination uniformly in $\cos\iota$ with 
$\iota\in[0^{\circ},30^{\circ}]$.

For each HEN trigger, the injections are distributed uniformly in time 
over the on-source window.  The injection sky positions are selected 
randomly following the estimated probability distribution 
(\ref{eq:lognormal}) for the HEN trigger, to account for the uncertainty in 
the true HEN direction of incidence.  The polarization angle 
(orientation of the rotational axis on the sky) is distributed 
uniformly.
Finally, the amplitude and arrival time at each detector is perturbed 
randomly to simulate the effect of calibration errors in the \lo~and 
\vo~detectors. 

\subsubsection{Exclusion distances}

For each waveform type we set a 90\% confidence level lower limit 
on the distance to a GW source associated with a given HEN 
trigger\footnote{Upper limits for each waveform and HEN 
trigger are available at 
\url{https://dcc.ligo.org/cgi-bin/DocDB/ShowDocument?docid=p1200006}.}.
This is defined as the maximum distance $D_{90\%}$ such that
for any distance $D \le D_{90\%}$ there is a probability of at least
0.9 that such a GW signal would have produced an event louder than the
loudest on-source event actually measured.
For inspirals, each distance corresponds to a well-defined amplitude.  We
can associate an amplitude to each distance for the
sine-Gaussian waveforms as well, by assuming a fixed energy in
gravitational waves.  For concreteness, we select $E_\textrm{GW} =
10^{-2} M_\odot c^2$.  This corresponds to the optimistic limit
of possible gravitational-wave emission by various processes in the
collapsing cores of rapidly rotating massive stars
(\citealt{fryer:02,kobayashi:03,Piro:2006ja,fryernew:11}, and 
discussion in Sec.~\ref{sec:sources}); more conservative
estimates based on simulations have been made in 
\citet{dimm,ott:09,scheidegger:10b,ott:11,tako}.  

\begin{figure}[!h]
\includegraphics[width=0.48\textwidth]{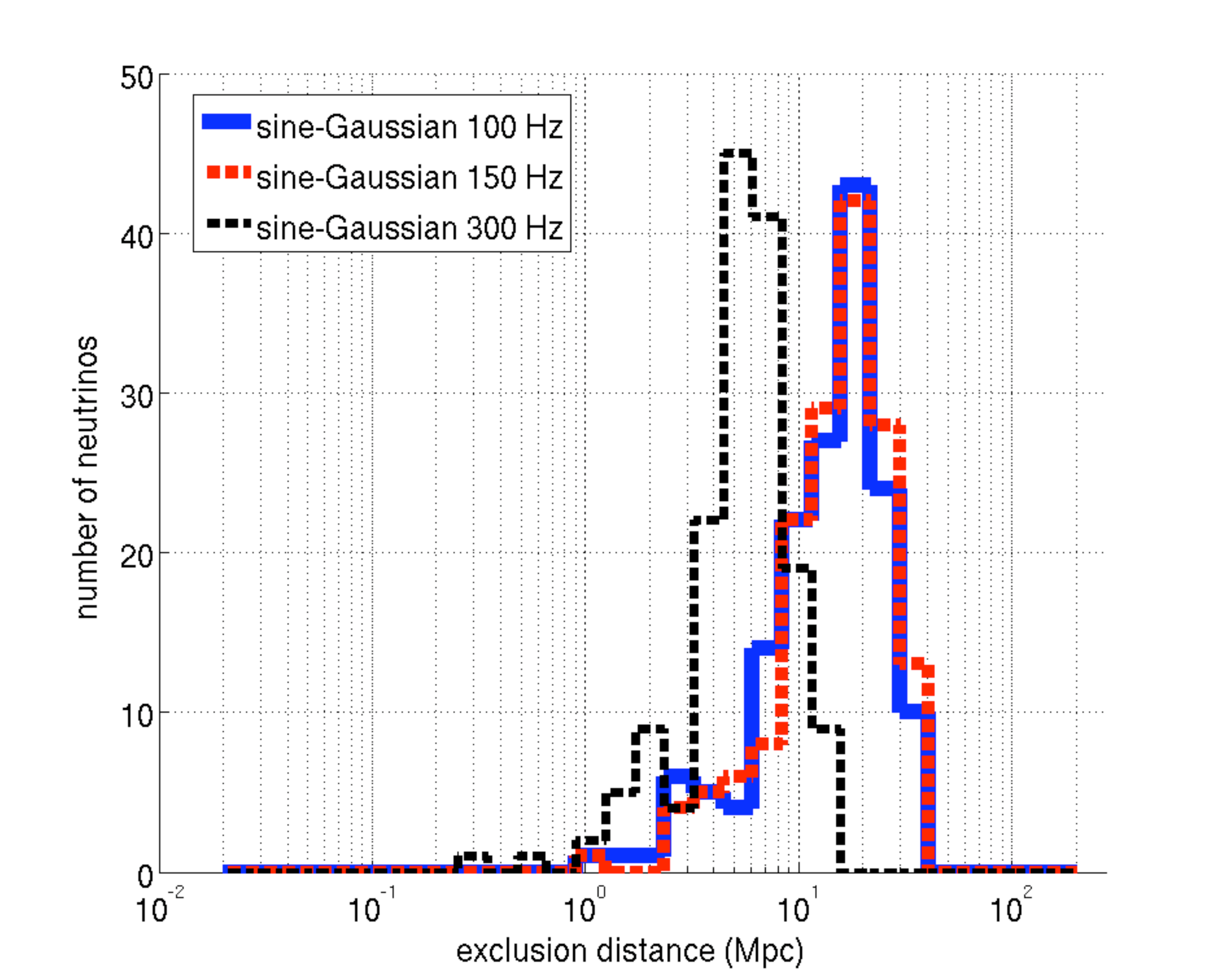}
\includegraphics[width=0.48\textwidth]{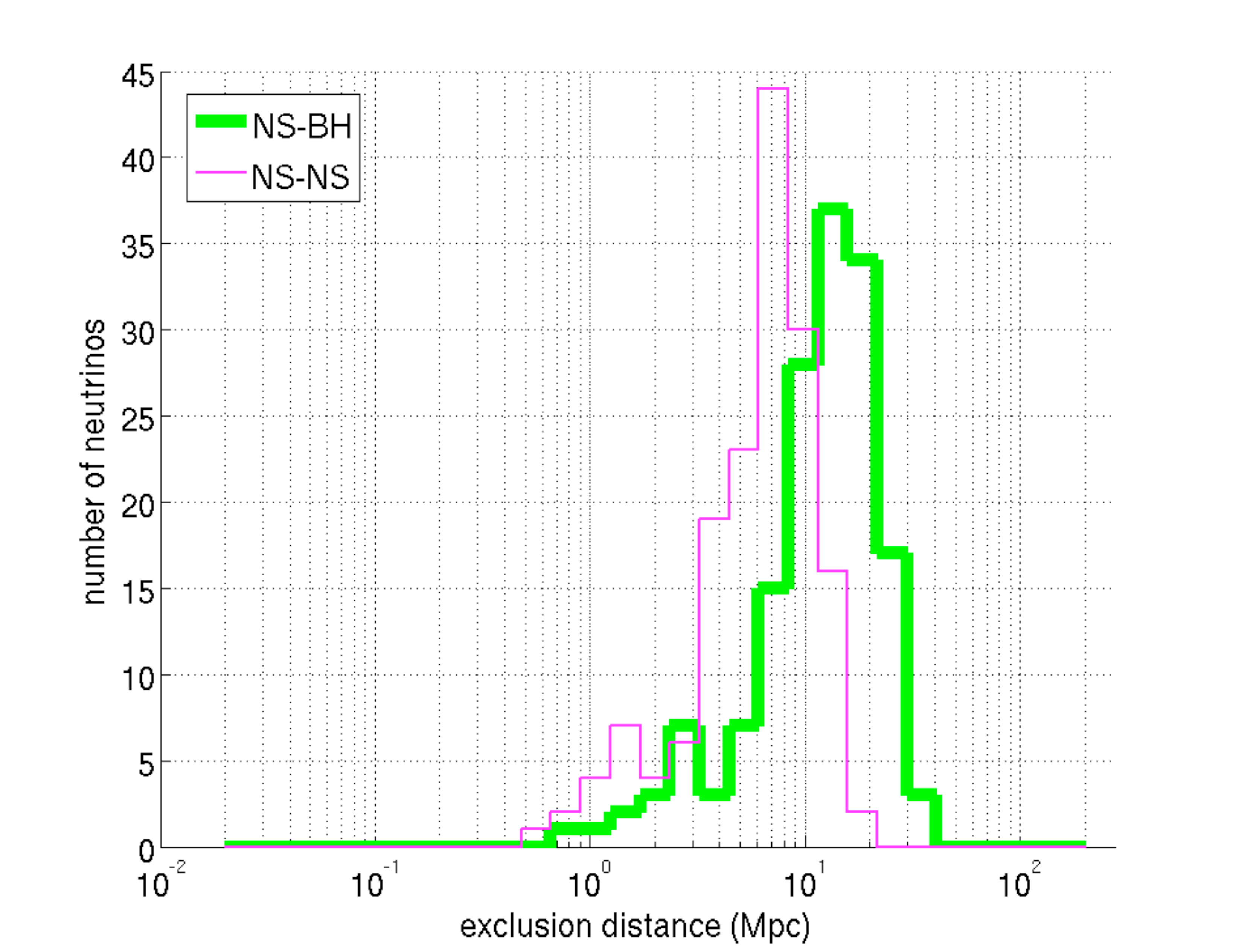}
\caption{Low-frequency analysis: the top plot is the histogram for the sample of analysed neutrinos of the distance exclusions at the $90\%$ confidence level for 
the 3 types of sine-Gaussian models considered: 100\,Hz, 150\,Hz and 300\,Hz. A standard siren gravitational wave emission of $E_{GW} = 10^{-2}\,M_{\sun}\,c^2$ is assumed.
The bottom plot shows the distance exclusions 
for the 2 families of binary inspiral models considered: NS-NS and BH-NS.} 
\label{fig:distances}
\end{figure}

\begin{figure}[h!]
\begin{center}
\includegraphics[width=0.48\textwidth]{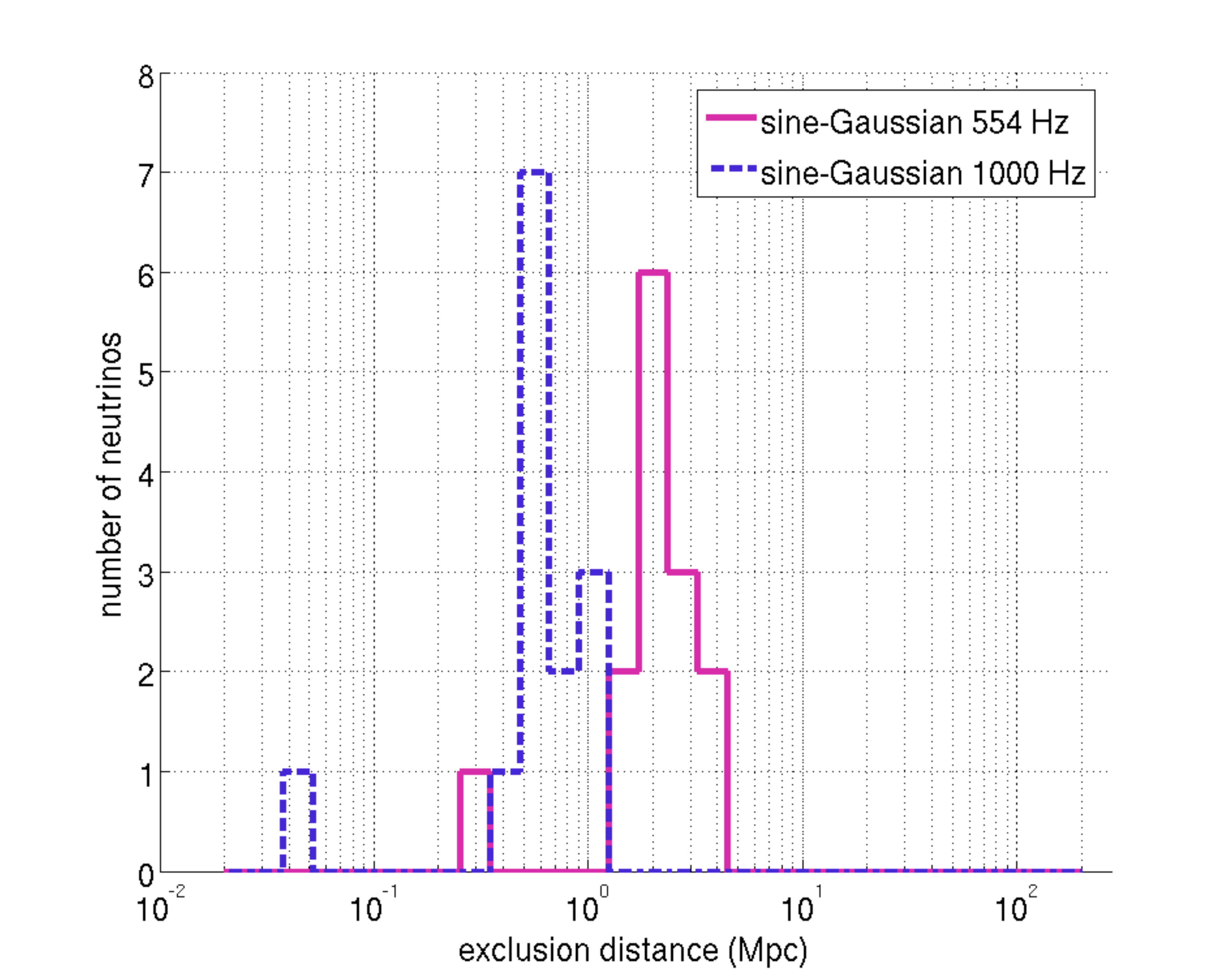}
\caption{High-frequency analysis: the histogram for the sample of analysed neutrinos of the distance exclusions at the $90\%$ confidence level for 
the 2 frequencies of circular sine-Gaussian models considered: 554\,Hz and 1000\,Hz.}
\label{fig:distances_2}
\end{center}
\end{figure}

For each type of gravitational wave simulated, the distributions of 
exclusion distances for our neutrino sample are shown in 
Figures~\ref{fig:distances} and~\ref{fig:distances_2}.
For binary neutron star systems of $(1.35-1.35) M_\odot$ and 
black hole - neutron star systems of $(5-1.35) M_\odot$ typical 
distance limits are 5\,Mpc and 10\,Mpc respectively.
For the sine-Gaussian waveforms 
with $E_\textrm{GW} = 10^{-2} M_\odot c^2$
we find typical distance limits between 5\,Mpc and 17\,Mpc
in the low-frequency band 
and of order 1\,Mpc in the high-frequency band.  
For other $E_\textrm{GW}$ the limits scale as  
$D_{90\%} \propto (E_\textrm{GW}/10^{-2} M_{\sun} c^2)^{1/2}$.
For example, for $E_\textrm{GW} = 10^{-8} M_\odot c^2$ (typical of 
core-collapse supernovae) a signal would only be observable from a Galactic source.

\section{Astrophysical implications}
\label{sec:astrophysics}

Observational constraints on joint sources of GW and HEN signals have been derived in \cite{BartosObs}. However, 
they are based on the interpretation and  the combination of previously published and independent GW and HEN observational results.
The results presented in this section are the first derived from a joint GW-HEN analysis, using concomitant data obtained with \lo/\vo~and \ant.

\subsection{Upper limits on GW-HEN populations}

The present search for GW and HEN correlations in space and time 
revealed no evidence for coincident events.  This implies a 
90\% confidence level upper limit on the rate of 
detectable coincidences of $2.3/T_{\textrm{obs}}$, where 
$T_{\textrm{obs}} \approx 90$ days is the duration of coincident 
observations. 
This can be expressed as a limit on the rate density (number 
per unit time per unit volume) $\rho_{\textrm{GW-HEN}}$ of 
joint GW-HEN sources:
\begin{equation}
\rho_{\textrm{GW-HEN}} \le \frac{2.3 {\cal F}_b}{V T_{\textrm{obs}}} \, .
\label{eq_astro1}
\end{equation}
Here ${\cal F}_b$ is the beaming factor (the ratio of the total number of 
sources to the number with jets oriented towards Earth\footnote{
For example, for a jet opening angle of $5^\circ$ gives 
${\cal F}_b\sim300$, while $30^\circ$ gives ${\cal F}_b\sim 10$.
}), and $V$ is the volume of universe probed 
by the present analysis for typical GW-HEN sources. 

We take as fiducial sources 
two classes of objects: the final 
merger phase of the coalescence of two compact objects (short GRB-like), 
or the collapse of a massive object (long GRB-like), both 
followed by the emission of a relativistic hadronic jet. 
We define the HEN horizon as the distance for which the probability to
detect at least 1 HEN in \ant~with 5 lines is 50\%.
In the case of short GRBs (SGRBs), the HEN horizon is estimated 
to be $d_{50}=4$\,Mpc using \cite{guetta:04}, while the typical GW horizon from the 
inspiral model is 5\,Mpc to 10\,Mpc depending on the binary masses.   
For long GRBs (LGRB) the HEN horizon increases to $d_{50}=12$\,Mpc using \cite{guetta:04}. 
The GW emission associated with long GRBs is highly uncertain; our 
optimistic assumption of $E_\textrm{GW}=10^{-2}M_\odot c^2$ at low 
frequencies gives a typical horizon distance of 10\,Mpc to 20\,Mpc in GW. 
Using the lower of the GW and HEN distances in each case yields 
from equation (\ref{eq_astro1}) approximate limits on the population 
density. For SGRB-like sources, related to the merger of two compact 
objects, we find 
$\rho^{\textrm{SGRB}}_{\textrm{GW-HEN}} \lesssim {\cal F}_b \times 10^{-2}\,\Mpc^{-3}\,\yr^{-1}$.
For LGRB-like sources, related to the collapse of massive stars, we 
find 
$\rho^{\textrm{LGRB}}_{\textrm{GW-HEN}} \lesssim {\cal F}_b E_{0.01}^{-3/2} \times 10^{-3}\,\Mpc^{-3}\,\yr^{-1}$, 
where $E_{0.01} \equiv E_\textrm{GW}/10^{-2} M_{\sun} c^2$.\\

\subsection{Comparison of limits with existing estimates}
\label{sec:comparisons}

\citet{guetta06}, \citet{nakar:06}, and \citet{guetta09} suggest a local 
rate density of SGRBs of 
$\rho_{\textrm{SGRB}} \sim 10^{-7}\,\Mpc^{-3}\,\yr^{-1}$ to $10^{-6}\,\Mpc^{-3}\,\yr^{-1}$ after correcting for beaming effects. 
This is similar to the abundance of binary neutron star mergers, their assumed  
progenitors, estimated to be 
$\rho_{\textrm{NS-NS}} \sim 10^{-8}\,\Mpc^{-3}\,\yr^{-1}$ to $10^{-5}\,\Mpc^{-3}\,\yr^{-1}$  
\citep[see for example][]{2010CQGra..27q3001A}, 
and well below the reach of the present search 
($\rho^{\textrm{SGRB}}_{\textrm{GW-HEN}} \lesssim {\cal F}_b\times10^{-2}\,\Mpc^{-3}\,\yr^{-1}$). 
With $T_{\textrm{obs}} = 1\,\yr$, an improvement of a factor 10 on the 
detection distance is required in order to begin constraining the 
fraction of mergers producing coincident GW$-$HEN signals.

\cite{guetta:05} estimate a total rate of long GRBs of 
$\rho_{\textrm{LGRB}} \sim 3\times10^{-8}\,\Mpc^{-3}\,\yr^{-1}$ after correcting for beaming effects; these 
sources are closely related to Type II and Type Ibc core-collapse 
supernovae. The local rate of SNIbc is 
$\rho_{\textrm{SNIbc}} \sim 2\times10^{-5}\,\Mpc^{-3}\,\yr^{-1}$  
\citep{Guetta2}, whereas 
$\rho_{\textrm{SNII}} \sim 2\times10^{-4}\,\Mpc^{-3}\,\yr^{-1}$  
\citep{Bazin}, relatively close to the obtained limit 
$\rho^{\textrm{LGRB}}_{\textrm{GW-HEN}} \lesssim {\cal F}_b E_{0.01}^{-3/2}\times10^{-3}\,\Mpc^{-3}\,\yr^{-1}$ 
under our optimistic assumptions of GW emission in this scenario.
A factor 10 only is required in order to begin constraining the 
fraction of stellar collapse events producing coincident weakly beamed GW-HEN signals, 
which translates into a required improvement of 2 on the detection distance.

\section{Conclusions}
\label{sec:conclusions}

This first joint GW-HEN search using 2007 data, obtained with
the \ant~HEN telescope and the \vo/\lo~GW interferometers,
opens the way to a novel multi-messenger astronomy.
Limits on the rate density 
$\rho_{\textrm{GW-HEN}}$ of joint GW-HEN emitting systems were extracted for the first time using
the analysis of coincident GW-HEN data. We note that these limits are consistent with the ones obtained in \cite{BartosObs} derived 
from independent GW-HEN observations.
More stringent limits will be available by performing similar coincidence 
analyses using other data sets provided by the same 
instruments.  

For instance, the sixth \lo~science run S6 and second and third \vo~science runs VSR2,3 covered the period from 7 July 2009 to 
21 October 2010.
Meanwhile, the \ant~telescope has taken data with first 10 then 12 active 
lines since the end of December 2007.  Their enhanced sensitivities should 
permit a combined analysis to gain the factor required to obtain 
$\rho^{\textrm{LGRB}}_{\textrm{GW-HEN}} \le \rho_{\textrm{SNII/SNIbc}}$ and 
begin to constrain the fraction of stellar collapse events accompanied by 
the coincident emission of relativistic jets beamed towards Earth. 
The analysis of these data is underway, and a similar analysis using data 
from the \lo/\vo~S5-VSR1 periods and the \ice~HEN telescope in its 22 string 
configuration is being finalized.

Future observing runs involving IceCube, \km~\citep{Km3}, 
and the advanced \lo~and advanced \vo~projects \citep{AdvancedITF}, are likely 
to coincide as well. They will give other opportunities to look for 
potential coincident GW-HEN emissions. 

\section{Acknowledgments}
The authors gratefully acknowledge the support of the United States National Science Foundation for the construction and operation of the
\lo~Laboratory, the Science and Technology Facilities Council of the United Kingdom, the Max-Planck-Society, and the State of
Niedersachsen/Germany for support of the construction and operation of the GEO600 detector, and the Italian Istituto Nazionale di Fisica
Nucleare and the French Centre National de la Recherche Scientifique for the construction and operation of the \vo~detector. The authors
also gratefully acknowledge the support of the research by these agencies and by the Australian Research Council, 
the International Science Linkages program of the Commonwealth of Australia, the Council of Scientific and Industrial Research of India, 
the Istituto Nazionale di Fisica Nucleare of Italy, the Spanish Ministerio de Educaci\'on y Ciencia, the Conselleria d'Economia Hisenda i Innovaci\'o of the
Govern de les Illes Balears, the Foundation for Fundamental Research on Matter supported by the Netherlands Organisation for Scientific Research, 
the Polish Ministry of Science and Higher Education, the FOCUS Programme of Foundation for Polish Science, the Royal Society, the Scottish Funding Council, the
Scottish Universities Physics Alliance, The National Aeronautics and Space Administration, the Carnegie Trust, the Leverhulme Trust, the
David and Lucile Packard Foundation, the Research Corporation, and the Alfred P. Sloan Foundation.

The authors also acknowledge the financial support of the funding agencies for the construction and operation of the {\ant} 
neutrino telescope: Centre National de la Recherche Scientifique (CNRS), Commissariat \`{a} l'\'energie atomique et aux \'energies alternatives (CEA), 
Agence National de la Recherche (ANR), Commission Europ\'eenne (FEDER fund and Marie Curie Program), R\'egion
Alsace (contrat CPER), R\'egion Provence-Alpes-Cote d'Azur, D\'epartement du Var and Ville de La Seyne-sur-Mer, France; Bundesministerium 
f\"{u}r Bildung und Forschung (BMBF), Germany; Istituto Nazionale di Fisica Nucleare (INFN), Italy; Stichting voor
Fundamenteel Onderzoek der Materie (FOM), Nederlandse organisatie voor Wetenschappelijk Onderzoek (NWO), the Netherlands;
Council of the President of the Russian Federation for young scientists and leading scientific schools supporting grants, Russia; National 
Authority for Scientific Research (ANCS), Romania; Ministerio de Ciencia e Innovacion (MICINN), Prometeo of Generalitat
Valenciana (GVA) and Multi-Dark, Spain. They also acknowledge the technical support of Ifremer, AIM and Foselev Marine for
the sea operation and the CC-IN2P3 for the computing facilities. This publication has been assigned {\lo}~Document Number LIGO-{\dccnumber}.

\bibliographystyle{apj} 

\end{document}